\documentclass[11pt,a4paper]{article} 
\pdfoutput=1
\usepackage{jheppub}
\usepackage{amsfonts}
\usepackage{mathtools}
\usepackage{dcolumn}	
\usepackage{slashed}	
\usepackage{bm}		
\usepackage{dsfont} 		
\usepackage[T1]{fontenc}
\usepackage{textcomp}
\usepackage{ragged2e}
\newcommand{\up}[1]{\textsuperscript{#1}}		
\newcommand{\tabref}[1]{table \ref{tab:#1}}		
\newcommand{\figref}[1]{figure \ref{fig:#1}}		
\newcommand{\citeR}[2][]{ref{#1}.~\cite{#2}}		
\renewcommand{\eqref}[1]{eq.~(\ref{eq:#1})}		
\newcommand{\eten}[1]{\ensuremath{\times 10^{#1}}}			
\newcommand{\Tr}[1]{\ensuremath{\text{Tr}\left[ #1 \right]}	}	
\newcommand{\lb}{\ensuremath{\left}}		
\newcommand{\rb}{\ensuremath{\right}}

\newcommand{\MS}{\ensuremath{\overline{\text{MS}}} }

\newcommand{\spartial}{\ensuremath \slashed{\partial}}

\newcommand{\lrD}[1]{\ensuremath{\!\! \stackrel{\leftrightarrow}{D_{#1}}\!\!}}
\newcommand{\nl}{\nonumber \\ & \quad }

\begin{document}

\title{Probing the fermionic Higgs portal at lepton colliders}
\author[a,b]{Michael A.~Fedderke,}
\author[b]{Tongyan Lin,}
\author[a,b]{and Lian-Tao Wang}

\affiliation[a]{Department of Physics, The University of Chicago, Chicago, Illinois, 60637, USA}
\affiliation[b]{Enrico Fermi Institute and Kavli Institute for Cosmological Physics, The University of Chicago, Chicago, Illinois, 60637, USA}

\emailAdd{mfedderke@uchicago.edu}
\emailAdd{liantaow@uchicago.edu}
\emailAdd{tongylin@gmail.com}

\keywords{Higgs Physics, Effective Field Theories}

\arxivnumber{1506.05465}

\abstract{
We study the sensitivity of future electron-positron colliders to UV completions of the fermionic Higgs portal operator $H^\dagger H \bar \chi \chi$. Measurements of precision electroweak $S$ and $T$ parameters and the $e^+e^- \to Zh$ cross-section at the CEPC, FCC-ee, and ILC are considered. The scalar completion of the fermionic Higgs portal is closely related to the scalar Higgs portal, and we summarize existing results. 
We devote the bulk of our analysis to a singlet-doublet fermion completion. Assuming the doublet is sufficiently heavy, we construct the effective field theory (EFT) at dimension-6 in order to compute contributions to the observables. We also provide full one-loop results for $S$ and $T$ in the general mass parameter space. In both completions, future precision measurements can probe the new states at the (multi-)TeV scale, beyond the direct reach of the LHC.
}

\maketitle
\flushbottom

\section{Introduction}
Among the particularly compelling scenarios in which new physics couples to the Standard Model (SM) are the so-called vector, neutrino and Higgs ``portals,'' which involve the three lowest-dimension SM-singlet operators which can couple to new physics (see, e.g., \citeR{Batell:2009di} and references therein). Of these, the Higgs portal, which encompasses operators of the form $H^\dagger H {\cal O}_{\text{NP}}$ where ${\cal O}_{\text{NP}}$ is a SM-singlet operator built from new fields, has been the focus of much interest in the literature (e.g., \citeR[s]{McDonald:1993ex,Burgess:2000yq,Patt:2006fw,Kim:2008pp,Englert:2011yb,Baek:2011aa,Djouadi:2011aa,Batell:2011pz,Greljo:2013wja,Chacko:2013lna,Cline:2013gha,Walker:2013hka,Robens:2015gla,Freitas:2015hsa}), especially in the wake of the 2012 discovery \cite{Aad:2012tfa,Chatrchyan:2012ufa} of a light Higgs boson with SM-like properties \cite{ATLAS-CONF-2015-007,Khachatryan:2014jba}. There are a number of reasons for its appeal aside from its low dimensionality. The Higgs portal provides a possible coupling of the dark matter (sector) to the SM, as we discuss in more detail below. Some well-studied models of this class---such as $H^\dagger H S^2$ with $S$ a scalar singlet---have the ability to effect non-trivial modification of the dynamics of electroweak symmetry breaking (EWSB), rendering a first-order phase transition possible, with interesting implications for electroweak baryogenesis~\cite{Espinosa:2007qk,Gonderinger:2009jp,Chung:2012vg,Fairbairn:2013uta,Li:2014wia,Profumo:2014opa}. Additionally, such models may characterize the low-energy physics in theories which ameliorate the SM hierarchy problem \cite{Craig:2013xia}.

\begin{figure}[t]
\centering
\includegraphics[width=0.75\textwidth]{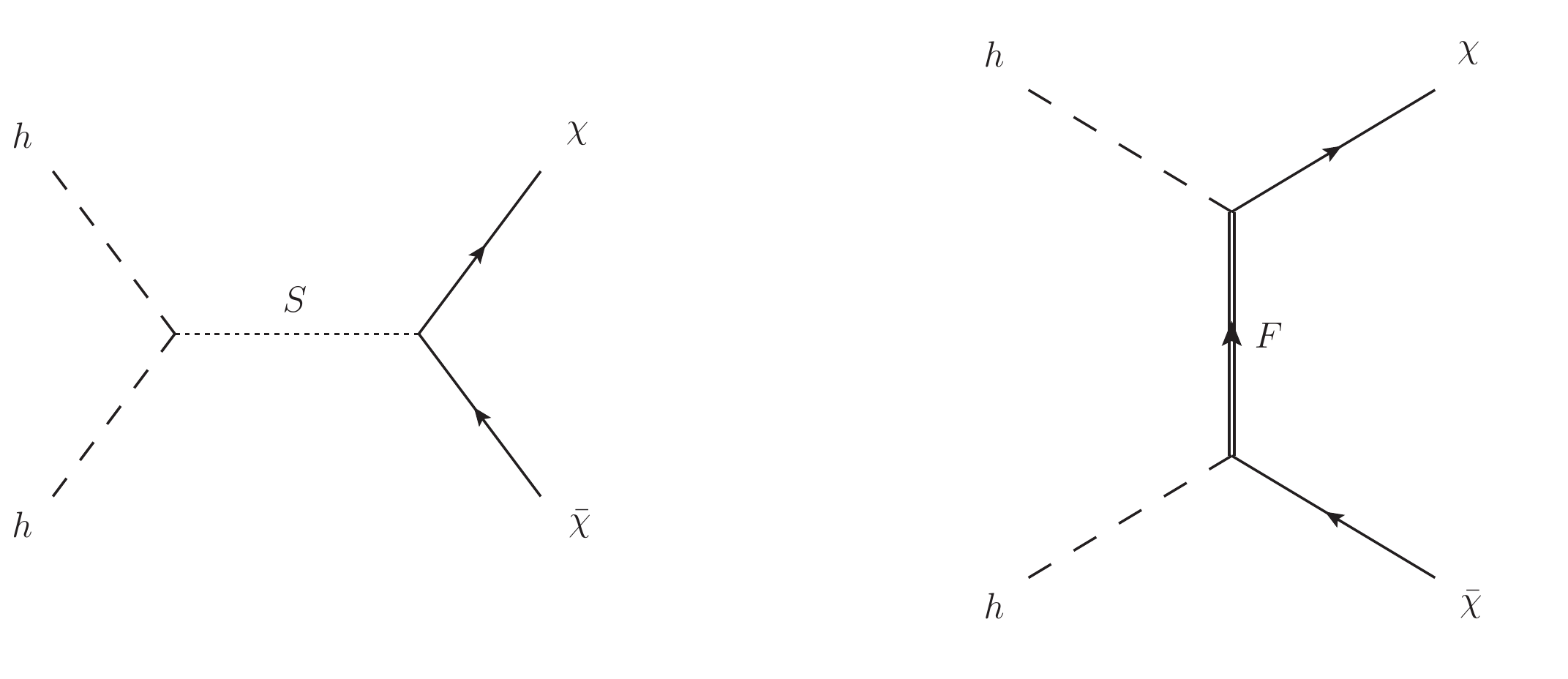}
\caption{ \label{fig:completions} Two possible UV completions of the fermionic Higgs portal operator, $H^\dagger H \bar{\chi}\chi$.  In the ``scalar completion'' (left diagram), a singlet scalar $S$ has a renormalisable coupling to the $H$ doublet and a Yukawa coupling to the singlet $\chi$ field; in the ``fermionic completion'' (right diagram), a Yukawa-like coupling exists between a vector-like $SU(2)$ doublet $F$ and the $H$ and $\chi$ fields.}
\end{figure}

While the simple picture of only considering the lowest-dimension fermionic Higgs portal operator in a bottom-up analysis is of course instructive, much more can be said about particular possible UV completions of this model. In this paper we study two such completions (see \figref{completions}) for the CP-even fermionic Higgs portal operator, $H^\dagger H \bar{\chi}\chi $, and their precision constraints. In the first model, a heavy singlet $S$ acts as a mediator, with a Yukawa coupling to the fermion $\chi$ (which we take to be Dirac) and renormalisable couplings to the Higgs-sector of the SM. This is essentially an extension of the singlet-scalar Higgs portal scenario. In the second model, we introduce a new heavy vector-like $SU(2)$ fermion doublet $F$ and couple both the singlet and doublet fermions to the SM Higgs: $-\kappa H \bar{F} \chi + \text{h.c.}$. Such singlet-doublet models have also been studied in the past~\cite{Enberg:2007rp,Cynolter:2008ea,Cohen:2011ec}.

Since the new states in either completion are coupled to the Higgs-sector of the SM (and in the latter scenario, also have non-trivial charges under the SM gauge group), precision electroweak and Higgs physics will be important probes of these potential UV completions; for example, measurements of $S$, $T$ and the Higgsstrahlung cross-section $\sigma_{Zh}$ are particularly sensitive. Although the present electroweak precision constraints have been known for a long time to be stringent, the current round of conceptual studies for future precision high-energy $e^+e^-$ machines---e.g., the ILC, CEPC, and FCC-ee\footnote{The collider formerly known as TLEP.}---have resulted in significantly sharper projections for how the constraints could improve. This brings into even clearer focus the discovery and exclusion potential possible through a study of these probes. It is the major goal of this paper to more fully understand how the projected sensitivities of these machines will allow stringent constraints to be placed on the UV completions of the Higgs portal that we consider.

Although we do not impose a dark-matter interpretation on the singlet $\chi$ in our models, it is worth noting this  possibility has received much attention recently. For the dimension-five fermionic Higgs portal, $H^\dagger H ( a \bar{\chi}\chi + i b \bar{\chi} \gamma_5 \chi)$, which comprises both CP-even and CP-odd coupling terms, a thermal-relic dark-matter interpretation is tightly constrained in the CP-even scenario by spin-independent direct detection constraints, whereas for the CP-odd coupling the direct detection constraints are alleviated and much parameter space remains open (see, e.g., \citeR[s]{Djouadi:2011aa,Freitas:2015hsa,LopezHonorez:2012kv,Fedderke:2014wda,deSimone:2014pda} and references therein). Turning to the UV completion of these operators, additional parameter space for thermal-relic dark matter is available when the new mediator mass (either the doublet fermion or singlet scalar) is similar to or below the mass scale of $\chi$. For example, the relic density may be set by coannihilation of the charged and neutral fermions \cite{Cohen:2011ec,Yaguna:2015mva}, or by dark matter annihilation into lighter scalars (see \citeR[s]{Baek:2011aa,Baek:2012uj,Baek:2014jga,Khoze:2015sra,Baek:2015lna} for studies of the singlet scalar UV completion).  These scenarios can have small couplings and thus evade electroweak precision and other constraints. As our focus is not on the dark-matter interpretation of $\chi$, but rather on precision probes of the fermionic Higgs portal coupling and its UV completions generally, we restrict ourselves to the regime where the new mediator is heavy and the EFT is valid.

The remainder of this paper is structured as follows: we begin in section \ref{sect:sensitivities} summarizing the experimental sensitivities on measurements of $S$, $T$ and $\sigma_{Zh}$ which are currently projected to be attainable at future colliders. In section \ref{sect:singletscalar} we introduce and briefly discuss the singlet-scalar UV completion, summarizing existing results in the literature. In section \ref{sect:singletdoublet} we introduce the fermionic UV completion and discuss our general expectations. We then detail the construction of the EFT we used to analyse this model in section \ref{sect:singletdoubletEFT}, and give the EFT results for the precision electroweak and Higgsstrahlung constraints in sections \ref{sect:singletdoubletEWPO} and \ref{sect:singletdoubletHiggs}, respectively. In section \ref{sect:loops} we discuss the one-loop computation of the precision electroweak limits in the general mass parameter space for this model. We conclude in section \ref{sect:conclusions}. Appendix \ref{app:STloops} contains technical details on the one-loop computation of $S$ and $T$, while Appendix \ref{app:mix_match} contains a clarification on a technical point of the EFT matching.
 
\section{Experimental sensitivities}
\label{sect:sensitivities}

Several proposed $e^+e^-$ collider experiments which can make advances in precision electroweak and Higgs physics measurements are currently under consideration: the International Linear Collider (ILC)~\cite{Baer:2013cma}, the Future Circular Collider (FCC-ee)~\cite{Gomez-Ceballos:2013zzn} and the Circular Electron Positron Collider (CEPC)~\cite{preCDR}. We summarize the sensitivities for various present and future measurements of the electroweak $S$ and $T$ parameters~\cite{Peskin:1990zt,Peskin:1991sw}, and the Higgsstrahlung cross-section $\sigma_{Zh}$, along with the references on which they are based, in \tabref{sensitivities}.  For each collider, we include several scenarios for the run parameters in order to compare the improvements in sensitivity that are possible with collider upgrades.

\begin{table}
\centering
\caption{\label{tab:sensitivities} Experimental sensitivities for current and future measurements of $(S,T)$ and $\sigma_{Zh}$ assumed in this paper. The $\sigma_{Zh}$ sensitivities are quoted as 68\% confidence percentage uncertainties, while the $(S,T)$ limits are shown as the parameters $(\, \sigma_{\textsc{s}},\, \sigma_{\textsc{t}},\, \rho_{\textsc{st}}\, )$ which define the covariance matrix in \eqref{cov_matrix}, and are extracted from the 68\% coverage likelihood contours ($-2 \log[ {\cal L} / {\cal L}_0 ]=2.30$) in the indicated references.\up{$\dagger$} The individual ILC Higgsstrahlung constraints must be combined where necessary by summing the relevant $-2 \log[ {\cal L} / {\cal L}_0 ]$ values to obtain the constraints for the three scenarios we consider in this paper: 250/fb @ 250~GeV, 250/fb @ 250~GeV + 500/fb @ 500~GeV, and 1.15/ab @ 250~GeV + 1.6/ab @ 500~GeV.}
\begin{tabular}{lll}
\multicolumn{3}{c}{\textbf{Precision Electroweak $\bm{(S,T)}$}} \\ \hline
\textbf{Scenario}		&	$\bm{(\, \sigma_{\textsc{s}}\ ,\ \sigma_{\textsc{t}}\ ,\ \rho_{\textsc{st}}\, )}$	&	\textbf{Reference}	 \\ \hline
Current										& (\,8.62\eten{-2}\ ,\ 7.37\eten{-2}\ ,\ 0.906\,)	&\cite{Fan:2014vta}, figure 1\\
CEPC ``Baseline''								& (\,2.39\eten{-2}\ ,\ 1.93\eten{-2}\ ,\ 0.844\,)	&\cite{Fan:2014vta}, figure 4\\
CEPC ``Improved $\Gamma_Z, \ \sin^2\theta\,$''		& (\,1.14\eten{-2}\ ,\ 8.79\eten{-3}\ ,\ 0.518\,)	&\cite{Fan:2014vta}, figure 4\\
CEPC ``Improved $\Gamma_Z, \ \sin^2\theta, \ m_t$''	& (\,1.12\eten{-2}\ ,\ 7.26\eten{-3}\ ,\ 0.779\,)	&\cite{Fan:2014vta}, figure 4\\
ILC											& (\,1.71\eten{-2}\ ,\ 2.14\eten{-2}\ ,\ 0.891\,)	&\cite{Fan:2014vta}, figure 1\\
FCC-ee-Z (aka TLEP--Z)							& (\,9.32\eten{-3}\ ,\ 8.70\eten{-3}\ ,\ 0.440\,)	&\cite{Fan:2014vta}, figure 1\\
FCC-ee-t	 (aka TLEP--t)							& (\,9.24\eten{-3}\ ,\ 6.18\eten{-3}\ ,\ 0.794\,)	&\cite{Fan:2014vta}, figure 1\\ \hline \hline
\multicolumn{3}{c}{\textbf{Higgsstrahlung $\bm{\sigma_{Zh}}$}} \\ \hline
\textbf{Scenario}		&	$\bm{( \Delta \sigma_{Zh} ) / \sigma_{Zh}}$ \textbf{(\%)} 	&	\textbf{Reference}	 \\ \hline
CEPC: 5/ab @ 240~GeV							&	0.5\%	&\cite{preCDR} \\
ILC:	250/fb @ 250~GeV							& 	2.6\%	&\cite{Asner:2013psa,Dawson:2013bba}\\
ILC: 	500/fb @ 500 GeV							&	3.0\%	&\cite{Asner:2013psa,Dawson:2013bba}\\
ILC: 1.15/ab @ 250 GeV							&	1.2\%	&\cite{Asner:2013psa,Dawson:2013bba}\\
ILC: 1.6/ab @ 500 GeV							&	1.7\%	&\cite{Asner:2013psa,Dawson:2013bba}\\
FCC-ee: 10/ab	@ 240 GeV (4IP) 					&	0.4\%	&\cite{Gomez-Ceballos:2013zzn}
\end{tabular}
\begin{flushleft}
\justify
{\footnotesize \up{$\dagger$}In this parametrisation, the $T$-axis intercepts of the 68\% coverage likelihood contour for two parameters are at $S=0, \ T=\pm \, \sigma_{\textsc{t}} \big[1 - \rho_{\textsc{st}}^2  \big]^{1/2} \cdot \big[ (\Delta\chi^2)_2^{68\%} \big]^{1/2} $ where $(\Delta\chi^2)_2^{68\%} = 2.30$, and similarly for the $S$-axis intercepts. In particular, this means that for $S=0$, the coverage in the $T$ parameter is about a factor of 2 better when going from the FCC-ee-Z to the FCC-ee-t scenario, despite the modest 30\% decrease in $\sigma_{\textsc{t}}$.} 
\end{flushleft}
\end{table}

For the $S$ and $T$ electroweak precision obserable (EWPO) limits, we present a parametrisation of the covariance matrix used to construct the 68\% coverage likelihood contours presented graphically in the references. We made the simplifying assumption of exact Gaussian uncertainties centered at $(S,T)=(0,0)$, and have parametrised the covariance matrix as 
\begin{align}
\Sigma = \begin{pmatrix} 
		\sigma_{\textsc{s}}^2 								& \rho_{\textsc{st}}\, \sigma_{\textsc{s}}\, \sigma_{\textsc{t}} \\  
		 \rho_{\textsc{st}}\, \sigma_{\textsc{s}}\, \sigma_{\textsc{t}}		& \sigma_{\textsc{t}}^2
	 \end{pmatrix}, \label{eq:cov_matrix}
\end{align}
from which the likelihood function is given by $-2 \log[ {\cal L} / {\cal L}_0 ] = \Delta^T \Sigma^{-1} \Delta$, where $\Delta^T = (S,T)$. The parameters $(\sigma_{\textsc{s}}, \sigma_{\textsc{t}}, \rho_{\textsc{st}} )$ were obtained by least-squares fitting of the constraint equation $-2 \log[ {\cal L} / {\cal L}_0 ]=2.30$ to a large number of $(S,T)$ co-ordinates read from the relevant graphically-presented 68\% coverage ellipses.

The precision electroweak sensitivity is based primarily on $Z$-pole measurements at the various colliders. For the CEPC, the ``baseline'' is a $Z$ mass threshold scan. The ``improved'' scenarios consider the gains if the collider is upgraded to allow for improved measurements of $\Gamma_Z$ (with the resonant depolarization method for energy calibration), $\sin^2 \theta$, or possibly $m_t$ (from an ILC top threshold scan). The FCC-ee-Z sensitivities are for $Z$-pole measurements with polarised beams (denoted ``TLEP-Z'' in \citeR{Fan:2014vta}). The FCC-ee-t scenario includes $Z$-pole measurements with polarised beams along with threshold scans for $WW$ and $t\bar{t}$ (denoted ``TLEP-t'' in \citeR{Fan:2014vta}) and has improved sensitivity owing to more precise measurements of the $W$ and top masses.

As has recently been emphasized in \citeR[s]{Englert:2013tya,Craig:2013xia} (and investigated further in, e.g., \citeR[s]{Cao:2014rma,Katz:2014bha,Craig:2014una,Henning:2014gca,Beneke:2014sba}),  the ``Higgsstrahlung'' process $e^+ e^- \rightarrow Zh$ can be measured at the percent or sub-percent level at future colliders and is a sensitive probe of new physics.%
\interfootnotelinepenalty=1000
\footnote{See also \citeR[s]{Hagiwara:2000tk,Dawson:2002wc,Barger:2003rs,Biswal:2005fh,Kile:2007ts,Dutta:2008bh} for previous work on constraining anomalous Higgs couplings with the Higgsstrahlung cross-section.} %
\interfootnotelinepenalty=100
Among the various ways a new-physics model modifies the Higgsstrahlung cross-section, a modification to the wavefunction renormalization of the Higgs (i.e., a modification of the momentum-dependent part of the Higgs two-point function) may be induced, leading to a ``Higgs oblique'' correction.

The limits on $\Delta \sigma_{Zh} / \sigma_{Zh}$ are quoted in \tabref{sensitivities} as the 68\% percentage confidence bounds given in the various references.  We consider a CEPC projection with  5/ab of data at $\sqrt{s} = $ 240~GeV  \cite{preCDR} and an FCC-ee projection with the assumption that there are 4 interaction points with 10/ab of combined integrated luminosity collected at  $\sqrt{s} = 240$~GeV  \cite{Gomez-Ceballos:2013zzn}. Various configurations of proposed runs for the ILC have been analyzed in the literature \cite{Asner:2013psa,Dawson:2013bba,Peskin:2013xra}; we consider only the configurations where (a) 250/fb of data are collected at $\sqrt{s} = 250$~GeV, (b) 500/fb of data at $\sqrt{s} = 500$~GeV are added to the 250/fb of data collected at $\sqrt{s} = 250$~GeV, and (c) the integrated luminosity is increased to a total of 1.15/ab at $\sqrt{s} = 250$~GeV and 1.6/ab at $\sqrt{s} = 500$~GeV.

\section{Scalar UV completion}
\label{sect:singletscalar}
The simplest possible UV completion of the CP-even fermionic Higgs portal is to take the Standard Model (SM) augmented by a vector-like Dirac fermion SM-singlet $\chi$ and a SM-singlet scalar $S$. These couple via the following Lagrangian
\begin{align}
\mathcal{L} = {\mathcal L}_{\textsc{sm}} &+ i \bar{\chi}\slashed{\partial}\chi - m_\chi \bar{\chi} \chi \nl + \frac{1}{2} (\partial_\mu S)^2 - \frac{1}{2} m_S^2 S^2 - \frac{b}{3!} m_S S^3 - \frac{\lambda_S}{4!} S^4 
									+ a\, m_S S |H|^2 + \frac{\epsilon_S}{2}  S^2 |H|^2- \kappa_S S \bar{\chi}{\chi} , \label{eq:scalar_add_lag}
\end{align}
where $a,b,\epsilon_S,\kappa_S,m_S$ and $m_\chi$ are real parameters. To be explicit, our sign and normalisation conventions for the SM Higgs-sector are
\begin{align}
\mathcal{L}_{\textsc{sm}} \supset |D_\mu H|^2 + \mu^2 |H|^2 - \lambda |H|^4,
\end{align}
and our unitary-gauge normalization conventions are such that $H = ( 0 , (v+h)/\sqrt{2} )$, with $v\approx 246$~GeV.

This model is just the renormalisable scalar Higgs portal model (see, e.g., \citeR{Henning:2014gca} for detailed discussion closely related to the present work), augmented with the singlet-$\chi$ Yukawa coupling. If we consider the limit where $bm_S, \epsilon_S \ll a m_S, \kappa_S$, and further assume that $m_S \gg v, m_\chi$, the mediator $S$ can be integrated out at tree-level to give rise to the following effective field theory:
\begin{align}
\mathcal{L}_{\textsc{EFT}} \supset {\mathcal L}_{\textsc{sm}} &+ i \bar{\chi}\slashed{\partial}\chi - m_\chi \bar{\chi} \chi \nl
													 + \frac{1}{2} a^2 |H|^4 - \frac{a \kappa_S}{m_S} H^\dagger H \bar{\chi}\chi  +  \frac{a^2}{m_S^2} \frac{1}{2} ( \partial_\mu |H|^2 )^2 + \frac{1}{2} \frac{\kappa_S^2}{m_S^2} (\bar{\chi}\chi)^2 + \cdots, \label{eq:scalar_add_lag_EFT}
\end{align}
where $\, \cdots $ represents terms proportional to various powers of $\epsilon_S, b$ and neglected higher-dimensional operators. Allowing terms proportional to $\epsilon_S, b m_S$ leads, at dimension-6, to only one additional operator, $|H|^6$, which taken together with the ability to change the sign of the $|H|^4$ operator in the SM-Higgs Lagrangian and yet maintain a stable minimum to the Higgs potential, can have interesting implications for the order of the electroweak phase transition \cite{Grojean:2004xa,Henning:2014gca}.

For the purposes of the discussion here, the $|H|^4$ operator can simply be absorbed into an unobservable shift of the SM Higgs quartic-coupling: $\lambda \rightarrow \lambda + \frac{1}{2}a^2$. The fermionic Higgs portal operator $H^\dagger H \bar{\chi}\chi$ leads to a variety of effects (see, e.g., \citeR{Fedderke:2014wda} for a detailed analysis): upon EWSB, it contributes to the mass of the $\chi$ field, and allows both $h \bar{\chi}\chi$ and $h^2 \bar{\chi}\chi$ couplings. The former coupling allows for $\chi$-$\chi$ scattering via Higgs exchange, and there is a further scattering contribution from the $(\bar{\chi}\chi)^2$ operator.

It is possible in this scenario to have the $\chi$ field play the role of the stable dark matter (DM), saturating the relic density. If the mediator $S$ is assumed to be somewhat heavier than the $\chi$, an EFT analysis is applicable \cite{Fedderke:2014wda,LopezHonorez:2012kv}: the relic abundance requirement fixes $m_S / (a\kappa_S) \sim 0.5$~TeV, while satisfying the LUX direct detection bounds \cite{Akerib:2013tjd} requires that the $\chi$ be fairly massive, $m_\chi^{\text{physical}} \gtrsim 3$~TeV. These requirements are actually in tension with the assumptions of an EFT analysis as they demand a very large coupling $a\kappa_S \gtrsim 2\pi$ to avoid $m_\chi > m_S$ and the resulting issues with perturbative unitarity \cite{Busoni:2013lha,Busoni:2014haa,Busoni:2014sya} for the EFT description of the non-relativistic DM freeze-out. This essentially means that, with the finely-tuned exception of $m_\chi \sim m_h/2$ \cite{Fedderke:2014wda,LopezHonorez:2012kv}, the heavy-mediator scenario is ruled out here. On the other hand, the scenario with a light mediator, $m_S < m_\chi$, is still viable as the direct detection constraints can be avoided while maintaining the correct relic abundance \cite{LopezHonorez:2012kv}. Of course, if we were to abandon the DM interpretation for $\chi$, and allow it to decay sufficiently quickly via some suitable modification to this model, the couplings $a$ and/or $\kappa_S$ can be freely dialed down without overclosing the universe, thereby alleviating the direct detection constraints and allowing light $m_\chi$.

In the EFT description for this model, the operator $\frac{1}{2} ( \partial_\mu |H|^2 )^2$ leads to a ``Higgs oblique'' correction to the Higgstrahlung process; i.e., the operator induces a Higgs wavefunction renormalization $\delta Z_h = a^2 v^2/m_S^2$, which modifies the cross-section by $\Delta \sigma_{Zh} / \sigma_{Zh} = \delta Z_h$ \cite{Craig:2013xia,Craig:2014una}. With a sensitivity of $\Delta \sigma_{Zh} / \sigma_{Zh} \sim 0.5$\% (corresponding closely to the CEPC and FCC-ee sensitivities in \tabref{sensitivities}), values of $m_S / a \lesssim 2.5$~TeV could be ruled out at 95\% confidence, with 5-$\sigma$ discovery reach up to $m_S / a \sim 1.6$~TeV \cite{Craig:2014una}. (The limits on $m_S/a$ just quoted are slightly weaker than those in figure 1 of \citeR{Henning:2014gca}, wherein a combined fit to all projected ILC Higgs coupling measurements is considered.)

 In this model,  the electroweak precision observables $S$ and $T$ are generated by one-loop running of the operators between $\mu = m_S$ and $\mu = m_W$, leading to a mixing of the Higgs oblique operator with the operators responsible for giving the $S$ and $T$ parameters. This one-loop running has been computed in \citeR{Henning:2014gca}, which considers the UV completion above in the limit that $\kappa_S \rightarrow 0$ (and with no assumption on the sizes of $\epsilon_S, b$). The leading contributions to the $S$ and $T$ parameters are found to be log-enhanced, but linearly proportional to the high-scale ($\mu = m_S$) value of the Wilson coefficient of the Higgs oblique operator, and suppressed by a loop factor.

To summarize the results of \citeR{Henning:2014gca}, the electroweak precision constraints on the model are marginally weaker than those from the Higgsstrahlung measurement, if the most optimistic Higgsstrahlung precision (with FCC-ee or CEPC) is compared to the most optimistic electroweak precision constraints (with FCC-ee-t).  Current constraints on $S$ and $T$ exclude $m_S/a \lesssim 1.2$~TeV at 95$\%$ confidence.

In the discussion above, we have neglected the impact of the neutral $\chi$ field that is still present in the EFT given by \eqref{scalar_add_lag_EFT}. This objection notwithstanding, provided that $m_\chi \gg E^*$, where $E^*$ is the energy scale for some process of interest, one can also integrate the $\chi$ out at one-loop, in which case the Wilson coefficient of the $\frac{1}{2} ( \partial_\mu |H|^2 )^2$ operator will shift by a subdominant amount; here, the shift will be smaller by a factor of $\sim \kappa_S^2/16\pi^2$ than the leading contribution. Therefore turning on $\kappa_S$ as a non-zero weak coupling, and integrating the $\chi$ out at the scale $\mu = m_\chi$, will give a subdominant one-loop shift to $\sigma_{Zh}$, and higher-loop contributions to $S$ and $T$. The ratio $\sigma_{Zh}:S:T$ should not change significantly, thus roughly preserving the relative strengths of the limits when $\chi$ is added to the theory. The foregoing limits on $m_S/a$ are therefore expected to be fairly accurate. They arise as an effect caused by the mediator, and are fairly insensitive to the presence of the portal coupling \emph{per se}.

In summary, the precision electroweak and Higgsstrahlung constraints on this scenario are very similar to those of the model in which the $\chi$ is simply absent, and this case has already received extensive attention in the literature.

\section{Singlet-doublet UV completion }
\label{sect:singletdoublet}
We devote the bulk of this paper to the singlet-doublet UV completion of the fermionic Higgs portal operator. The model consists of the Standard Model augmented by the same vector-like Dirac fermion SM-singlet $\chi$ as before, as well as a vector-like Dirac fermion $SU(2)$-doublet $F = (C,N)$ transforming under the SM gauge group as $(\bm{1},\bm{2},+1/2)$. The coupling of these particles to the SM is taken to be
\begin{align}
\mathcal{L} = {\mathcal L}_{\textsc{sm}} &+ i \bar{\chi}\slashed{\partial}\chi - m_\chi \bar{\chi} \chi + i \bar{F} \slashed{D} F - M_F \bar{F} F  - \kappa \bar{F} H \chi - \kappa \bar{\chi} H^\dagger F, \label{eq:add_lag}
\end{align}
where $D_\mu \equiv \partial_\mu - i g{W_\mu}^a t^a - i g' B_\mu Y$, and $t^a = \tfrac{1}{2} \sigma^a$. Without loss of generality we may absorb the phase of $\kappa$ into the definition of $\chi$ and/or $F$, and so throughout we take $\kappa$ to be a non-negative real parameter. Without much modification (it would suffice, e.g., to replace $m_\chi \bar{\chi}\chi \rightarrow m_\chi \bar{\chi}_{\textsc{l}} \chi_{\textsc{r}} + \text{h.c.}$, with $\arg{m_\chi}\neq 0$) it would also be possible to generate the CP-odd Higgs portal operator $H^\dagger H \bar{\chi} i \gamma_5 \chi$; we do not, however, consider this case further, and take all parameters to be real.

In the parameter region where $M_F \gg m_\chi$, the heavy doublet $F$ can be integrated out at the scale $\mu = M_F$, and the leading correction to the SM Lagrangian is the fermionic Higgs portal operator. We will consider constraints on this UV completion both in the regime where $M_F \gg m_\chi$, constructing an EFT to analyze the low-energy effects, as well as in the more general mass parameter space. For general masses, this model does not necessarily provide the UV completion to the fermionic Higgs portal operator as it is defined here and our EFT is not valid, so we instead perform direct one-loop computations of relevant observables.

Before discussing the computations and results in detail, it is worth laying out our general expectations. The Yukawa-like coupling $\kappa$ between the $F,\chi$ and $H$ fields in \eqref{add_lag} is a hard breaking of the accidental global $SU(2)_V$ custodial symmetry \cite{Sikivie:1980hm} of the SM, and as such we expect fairly large corrections to the precision electroweak $T$ parameter.  Additionally, the mass-splitting (i.e., weak iso-spin breaking) in the $F$ doublet which arises from the mixing of the neutral fermions $N$ and $\chi$ after electroweak symmetry breaking (EWSB), leads us to expect that there will additionally be contributions to the electroweak $S$ parameter. 

Note that, although we do not consider such a case in this paper, custodial symmetry could be restored---or broken in a controlled fashion---by augmenting the field content with an additional positively-charged vector-like fermion $\psi \sim (\bm{1},\bm{1},+1)$ with the same Yukawa-like coupling to the $H$ and $F$ fields; the mass-splitting $|m_\psi - m_\chi|$ then controls the degree to which the symmetry is broken (being restored in the degenerate-mass limit). There would however be additional experimental handles in this case as the new direct coupling of the Higgs doublet to electrically-charged fermions would lead to one-loop corrections to the $h\rightarrow\gamma\gamma$ rate.

We also expect that, in the model defined by \eqref{add_lag}, the Higgs oblique correction will be generated (by closed $N$-$\chi$ loops on the Higgs propagator, in the full-theory picture), so that even if we were to tune away the large corrections to $T$ by, e.g., the method indicated in the previous paragraph, significant constraints would still remain on this model. To be clear, however, the Higgs oblique correction is by no means the only contribution to the shift in the Higgsstrahlung cross-section which is induced in this model; indeed, non-zero $S$ and $T$ parameters themselves will also generate a shift to $\sigma_{Zh}$ (see \citeR{Craig:2014una}), and we take all such shifts into account.

\subsection{Effective field theory approach}
\label{sect:singletdoubletEFT}
It is instructive to begin our analysis by integrating out the new field content, assuming it is heavy, and considering the low-energy effects of the dimension-6 operators that are generated. In many popular bases of gauge-invariant dimension-6 operators which extend the SM (see, e.g., \citeR[s]{Grzadkowski:2010es,Buchmuller:1985jz,Hagiwara:1993ck,Willenbrock:2014bja}), one can immediately read off the $S$ and $T$ parameters as the Wilson coefficients of certain operators. In the so-called ``HISZ'' basis \cite{Hagiwara:1993ck}, $S$ is simply proportional to the Wilson coefficient of the operator $H^\dagger \hat{B}^{\mu\nu} \hat{W}_{\mu\nu} H$, and $T$ is likewise proportional to the Wilson coefficient of the operator $|H^\dagger D_\mu H|^2$; the observable corrections to the momentum-dependent part of the Higgs two-point function would arise from the operators $|H^\dagger D_\mu H|^2$ and $( \partial_\mu |H|^2 )^2$ in this basis, but the full $\sigma_{Zh}$ corrections require significantly more work to obtain \cite{Craig:2014una}. 

In the following, we first integrate out the $F$ doublet at the ``high-scale'' $\mu = M_F$, performing the one-loop matching onto an effective theory with the Standard Model field content plus the singlet $\chi$. We then repeat the procedure, integrating out the $\chi$ field at the ``low-scale'' $\mu = m_\chi$ and matching at one-loop onto a new EFT with the SM field content only. We assume the mass hierarchy $M_F \gg m_\chi \gg v \sim m_W$, where $M_F \gg m_\chi$ is imposed to guarantee that the leading correction upon integrating out the $F$ is the CP-even fermionic Higgs portal operator; this is not a necessary assumption of the model itself. 

In performing the matching, we find it convenient to match onto the basis of dimension-6 operators with SM fields shown in  \tabref{high_scale_operators}.  This basis of operators is not one of the standard ones and in particular certain operators (e.g., ${\cal O}_{(2,4)}$) are redundant in the sense that they are related to other operators by SM classical equations of motion (EOM). The redundant operators can be eliminated by field re-definitions whose effect is exactly equivalent to making replacements using the classical EOM (see, e.g., \citeR[s]{Arzt:1993gz,Grzadkowski:2010es}); this is \emph{correct even at the quantum level} if we consistently neglect higher-dimension operator corrections. We note for now that the operators in \tabref{high_scale_operators} form a convenient and sufficient set to match onto; we will return below to the treatment of the various redundancies in order to transform into the operator bases most convenient for calculating the  $S$ and $T$ parameters, and $\sigma_{Zh}$.

\begingroup
\renewcommand{\arraystretch}{1.11}
\begin{table}[t]
\centering
\caption{ \label{tab:high_scale_operators} Operators appearing in the effective theory below the scale $M_F$ (see text for sign and normalisation conventions). For operators with a $\chi$ bilinear, we have performed tree-level matching at the high-scale $M_F$ up to and including dimension-10 operators in order to be able to compute corrections to fairly high order in $m_\chi/M_F$ when later matching at $\mu = m_\chi$, but omit these lengthy results here as we used only selected terms from them as necessary (see text).
Operators containing only SM content appear in the one-loop matching both at the high scale $\mu=M_F$ and at the low scale $\mu=m_\chi$. The naming convention for the operators with only $n$ Higgs doublets $H$ and $m$ (covariant) derivatives is ${\cal O}_{(n,m)}$; other operators are named on an ad-hoc basis. We define $\hat{W}_{\mu\nu} \equiv i g W^a_{\mu\nu} t^a$ and $\hat{B}_{\mu\nu} \equiv i g' B_{\mu\nu} Y$. Note that the operator ${\cal O}_{(2,4)}$ is non-standard in the literature and can be eliminated using \eqref{O24}.} 
\resizebox{\textwidth}{!}{
\begin{tabular}{rl|rl}

\multicolumn{4}{c}{\textbf{ Tree-level --- operators containing a $\bm{\chi}$ bilinear }} \\ \hline
\textbf{Name} 			&	\textbf{Operator} 	& \textbf{Name} 	&	\textbf{Operator} \\ \hline \hline
${\cal O}_5$ 			& $H^\dagger H \bar{\chi}\chi$ &
${\cal O}_{6A}$ 		& $\tfrac{1}{2} \bar{\chi}\gamma^\mu \chi\ i ( H^\dagger D_\mu H - \text{h.c.} ) $ \\
${\cal O}_{6B}$ 		& $\tfrac{1}{2} H^\dagger H \ i ( \bar{\chi} \spartial \chi - \text{h.c.} ) $ &
${\cal O}_{7A}$			& $\tfrac{1}{2} H^\dagger H (  \bar{\chi} \Box \chi + \text{h.c.} ) $ \\
${\cal O}_{7B}$			& $|D_\mu H|^2 \bar{\chi}\chi $ &
${\cal O}_{7C}$			& $\tfrac{1}{2} \, i \lb( ( D_{\mu} H )^{\dagger} D_{\nu}H - \text{h.c.} \rb) \bar{\chi} \sigma^{\mu\nu} \chi $ \\
${\cal O}_{7D}$			& $\tfrac{1}{2} \lb( H^\dagger D_\mu H - \text{h.c.}  \rb) ( \bar{\chi} \gamma_\mu\spartial \chi - \text{h.c.} ) $ \hspace{2cm}&
etc.  \\[1ex] \hline \hline
\multicolumn{4}{c}{ \textbf{ One-loop --- operators with only SM content }}	 \\ \hline
${\cal O}_{(2,2)}$ 		& $| D_\mu H|^2$					&
${\cal O}_{(2,0)}$		& $|H|^2$							\\
${\cal O}_{(4,0)}$		& $|H|^4$							&
${\cal O}_{(4,2),A}$		& $\tfrac{1}{2} ( \partial_\mu |H|^2 )^2$ 	\\
${\cal O}_{(4,2),B}$		& $| H^\dagger D_\mu H |^2$ 			&
${\cal O}_{(4,2),C}$		& $|H|^2 |D_\mu H|^2$ 				\\
${\cal O}_{(6,0)}$		& $|H|^6$							&
${\cal O}_{(2,4)}$		& $H^\dagger D_\mu D^\nu D_\nu D^\mu H + \text{h.c.}$ \\
${\cal O}_{(WW)}$		& $H^\dagger \hat{W}_{\mu\nu} \hat{W}^{\mu\nu} H$ &
${\cal O}_{(BB)}$		& $H^\dagger \hat{B}_{\mu\nu} \hat{B}^{\mu\nu} H$ \\
${\cal O}_{(BW)}$		& $H^\dagger \hat{B}_{\mu\nu} \hat{W}^{\mu\nu} H$ &
${\cal O}_{(DW)}$		& $\Tr{ [ D_\mu, \hat{W}_{\nu\rho} ] [ D^\mu, \hat{W}^{\nu\rho} ] }$ \\[0.5ex]
${\cal O}_{(DB)}$		& $-\tfrac{(g')^2}{2} ( \partial_\mu B_{\nu\rho} ) ( \partial^\mu B^{\nu\rho} )$ &
${\cal O}_{(W)}$		& $( D_\mu H )^\dagger \hat{W}^{\mu\nu} (D_\nu H)$ \\
${\cal O}_{(B)}$			& $( D_\mu H )^\dagger \hat{B}_{\mu\nu} (D_\nu H)$ &
${\cal O}_{(WWW)}$		& $\Tr{\hat{W}_{\mu\nu} \hat{W}^{\nu\sigma} {\hat{W}_{\sigma}}^{\;\; \mu} }$ 
\end{tabular}}
\end{table}
\endgroup
 
\newcommand{\mcmf}[1]{\, m_\chi^{#1} / M_F^{#1} }
\begingroup
\renewcommand{\arraystretch}{1.2}
\begin{table*}[t]
\centering
\caption{ \label{tab:high_scale_coefficients} Wilson coefficients of the operators appearing in \tabref{high_scale_operators}, evaluated at $\mu = M_F$, the ``high scale.'' Our sign convention is that the Lagrangian term containing any operator of dimension greater than 4 which is shown in \tabref{high_scale_operators} appears multiplied by its corresponding Wilson coefficient and the appropriate inverse power of the EFT cutoff-scale $M_F$; \emph{however}, for operators of dimension 4 or less, the sign appearing in front of the operator follows the SM conventions and any multiplying power of $M_F$ is absorbed into the Wilson coefficient (see discussion in text). These expressions assume a hierarchy of scales $M_F \gg m_\chi$, and are correct to tree-level for the $c_i$ and to one-loop for the $B_{(j)}$. We present results to fairly high order in $m_\chi/M_F$, although not all of these terms are numerically necessary to obtain our results.}
\begin{tabular}{rl|rl}
\multicolumn{4}{c}{\textbf{Tree-level --- operators containing a $\bm{\chi}$ bilinear}} \\ \hline
\textbf{Coefficient} 	&	\textbf{Value}  & \textbf{Coefficient} 	&	\textbf{Value}\\ \hline
$c_5$ 		& $+\kappa^2$  \hspace{2.3cm}	& 	$c_{6A}$ 		& $+\kappa^2$ \\
$c_{6B}$ 		& $+\kappa^2$  	&	$c_{7A}$		& $-\kappa^2$ \\
$c_{7B}$		& $+\kappa^2$ 		&	$c_{7C}$		& $-\kappa^2$ \\
$c_{7D}$		& $-\kappa^2$ 		&	etc. 
\end{tabular}\vspace{1ex}
\resizebox{\textwidth}{!}{
\begin{tabular}{rl}
\multicolumn{2}{c}{\textbf{One-loop --- operators with only SM content}} \\ \hline
\textbf{Coefficient} 	&	\textbf{Value} \\ \hline
$B_{(2,0)}$ 	&  $- (\kappa ^2 / 4 \pi ^2) \, M_F^2\, \big[ 1 + \mcmf{} +\mcmf{2} +\mcmf{3} +\mcmf{4}$ \\
			&  \phantom{$- (\kappa ^2 / 4 \pi ^2) \, M_F^2\, \big[ 1$}\;$+\mcmf{5} + \mcmf{6} + \cdots \big]$ \\
$B_{(2,2)}$ 	& $+(\kappa ^2 / 16 \pi ^2 ) \big[ 1 - 2 \mcmf{} - 4 \mcmf{2}- 10\mcmf{3} -15\mcmf{4} $\\
			& \phantom{$+(\kappa ^2 / 16 \pi ^2 ) \big[ 1 $}\;$- 24\mcmf{5} - 32\mcmf{6} + \cdots \big]$ \\
$B_{(4,0)}$	& $-( \kappa^4 / 8\pi^2 ) \lb[ 1 + 4\mcmf{} + 9\mcmf{2} + 16 \mcmf{3}  +25 \mcmf{4} +\cdots \rb]$ \\
$B_{(4,2),A}$	 & $-(\kappa^4 / 24\pi^2 ) \lb[ 5 + 4 \mcmf{} - 9 \mcmf{2} \rb]$ \\
$B_{(4,2),B}$	 & $-(\kappa^4 / 48\pi^2 ) \lb[ 5 - 2 \mcmf{}  - 15 \mcmf{2} + \cdots \rb]$ \\ 
$B_{(4,2),C}$	 & $-(\kappa^4 / 48\pi^2 ) \lb[ 1 - 22 \mcmf{} - 123 \mcmf{2} + \cdots \rb]$ \\
$B_{(6,0)}$	 & $-(\kappa^6 / 12\pi^2 ) \lb[ 1 + 12 \mcmf{} + 48 \mcmf{2} \cdots \rb]$ \\
$B_{(2,4)}$	 & $+ (\kappa^2 / 48\pi^2 ) \lb[ 1 - \mcmf{} - 3\mcmf{2} - 20 \mcmf{3} - 40 \mcmf{4} + \cdots \rb]$ \\
$B_{(WW)}$	 & $- (\kappa^2 / 144\pi^2 ) \lb[ 1 - 4\mcmf{} + 3\mcmf{2} + 10 \mcmf{3} + 68\mcmf{4} + \cdots \rb]$ \\
$B_{(BB)}$	 & $- (\kappa^2 / 144\pi^2 ) \lb[ 1 - 4\mcmf{} + 3\mcmf{2} + 10 \mcmf{3} + 68\mcmf{4} + \cdots \rb]$ \\
$B_{(BW)}$	 & $- (\kappa^2 / 72\pi^2 ) \lb[ 1 - 4\mcmf{} + 3\mcmf{2} + 10 \mcmf{3} + 68\mcmf{4} + \cdots \rb]$ \\
$B_{(DW)}$	 & $+ (1 / 240\pi^2 )$ \\
$B_{(DB)}$	 & $+ (1 / 240\pi^2 )$ \\
$B_{(W)}$	 & $+ (\kappa^2 / 72\pi^2 ) \lb[ 7 - 4 \mcmf{} - 12 \mcmf{2}  + \cdots \rb]$ \\
$B_{(B)}$	 & $+ (\kappa^2 / 72\pi^2 ) \lb[ 7 - 4 \mcmf{} - 12 \mcmf{2}  + \cdots \rb]$ \\
$B_{(WWW)}$	 & Operator generated, but coefficient not computed.\\
			 & \quad Irrelevant to the observables of interest.
\end{tabular}}
\end{table*}
\endgroup

Upon integrating out $F$, we find the operators and Wilson coefficients shown in tables \ref{tab:high_scale_operators} and \ref{tab:high_scale_coefficients}. The only operators of dimension 4 or less appearing in \tabref{high_scale_operators} -- ${\cal O}_{(2,0)}$, ${\cal O}_{(4,0)}$, and ${\cal O}_{(2,2)}$ -- already appear in the SM. Our convention for these operators is such that the Wilson coefficients $B_{(2,0)}$, $B_{(4,0)}$, and $B_{(2,2)}$ simply add positively to the coefficients already present for these operators in the SM: i.e., ${\cal L} \supset ( 1 + B_{(2,2)} ) {\cal O}_{(2,2)} + ( \mu^2 + B_{(2,0)} ) {\cal O}_{(2,0)} - ( \lambda + B_{(4,0)} ) {\cal O}_{(4,0)}$. On the other hand, all operators with dimension greater than 4 are taken to appear in the Lagrangian as ${\cal L} \supset c_i\, M_F^{4 - d_i}\, {\mathcal O}_i$ where $c_i$ is the Wilson coefficient and $d_i$ is the dimension of the operator ${\mathcal O}_i$.

We worked to tree-level for the Wilson coefficients for any operator containing a $\chi$-field bilinear, and to one-loop in obtaining the Wilson coefficients for operators containing only SM field content.  The rationale is that whenever the $\chi$ later appears in any diagram computed in a consistent one-loop matching to obtain a Wilson coefficient of a dimension-6 operator containing only SM content, it will appear as a closed $\chi$ loop, so only the tree-level $\chi$-SM couplings are required. 

The tree-level matching was performed by inverting the classical equation of motion for the heavy doublet to solve for $F$ in terms of $H$ and $\chi$, and expanding in powers of $D_\mu/M_F$ (some lower-dimensional results were checked diagrammatically). The one-loop matching was performed diagrammatically utilizing dimensional regularization and the \MS renormalisation scheme.  Calculations were done in the broken electroweak phase, and on the EFT-side of the one-loop matching computations we consistently included diagrams with one or more tree-level $\chi$-SM coupling(s) and a single closed $\chi$ loop. As a check, some one-loop computations were performed both in the non-mass-diagonal basis of new neutral fermions, $N$ and $\chi$, and in the mass-diagonalised basis.%
\footnote{Mixing-angle effects are necessarily small in the range of EFT validity, being suppressed by at least $\kappa v/M_F$ for $M_F\gg m_\chi, \kappa v$ (see \eqref{theta_defn}). It is well-known (see, e.g., \S 3.3 (pp.~37--38) of \citeR{Skiba:2010xn}, and references therein) that physical effects which in the full theory would be ascribed to these small mixing angles are instead captured in the EFT description through the inclusion of higher-order operators, whose Wilson coefficients have their genesis (in part) in the UV-theory mixing angles. Readers unfamiliar with this point may wish to consult Appendix \ref{app:mix_match} for further comments.} %
In the latter case, expansion of the result in powers of the Higgs vev $v$ allows the comparison of these two methods and the extraction of the Wilson coefficients. %
Finally, we have checked that, to the order we work, the EFT and full theory one-loop computations reproduce the same IR-divergent%
\footnote{In the sense that they diverge if $m_\chi \rightarrow 0$.} %
logarithms ($\sim \log\lb[ m_\chi^2 / M_F^2 \rb]$ at $\mu = M_F$) as is of course required.

The EFT power-counting we employed is such that we kept only leading terms in $v/M_F$, but retained terms at higher order in $m_\chi / M_F$, although this was not always numerically necessary to obtain the results we present below. In order to compute these higher-order terms in the mass-ratio $m_\chi/M_F$ in our one-loop matching, we have systematically carried out the tree-level matching up to and including dimension-10 operators so as to capture the operators with higher numbers of derivatives acting on the $\chi$ field since, in dimensional regularization and \MS\!\!, such higher-derivative operators contribute positive powers of $m_\chi$ to the $S$-matrix element when appearing in conjunction with a closed $\chi$-loop. This latter point is perhaps opaque, and is illuminated by a simple concrete example: the term in $B_{(WW)}$ suppressed by $m_\chi^4/M_F^{4}$ relative to the leading term can be found from, e.g., the $v^2 (p^2/M_F^2)(m_\chi^4/M_F^4)$ part of the $ZZ$ two-point function at one-loop (along with other $VV$ two-point functions so as to break the degeneracy between $B_{(WW)}, B_{(BW)},$ and $B_{(BB)}$). Simple power-counting then indicates that, owing to the $M_F^{-6}$ suppression, this requires in turn knowledge of the subset of the operators at dimension-10 in the tree-level matching which contain two $Z$ boson fields, two Higgs fields, one $\bar{\chi}$ field, one $\chi$ field, two derivatives acting on the $Z$ fields, and--to be dimensionally correct--one derivative acting on either of the $\chi$ or $\bar{\chi}$ fields. When the $\chi$ loop is closed and the Higgs field replaced by its vev, such operators contribute to the $ZZ$ two-point function in the EFT with a term proportional to $(v^2p^2/M_F^6) \int d^dk \, k^2 / (k^2-m_\chi^2) \sim (v^2p^2/M_F^6)m_\chi^4$, as desired.

Having obtained the mixed tree-level/one-loop matching conditions, one should by all rights run the theory down from the high-scale $\mu = M_F$ to the ``low-scale'' $\mu = m_\chi$, integrate the $\chi$ out at this scale, and then run the theory to the scale required to consider any process of interest (i.e., the matching conditions supply only the initial conditions for the renormalisation group equation (RGE) running). However, we will choose to not run the operators using the RGE. The errors one makes in ignoring the running between $\mu=M_F$ and $\mu=m_\chi$ are proportional to $\log\lb[ m_\chi^2 / M_F^2 \rb]$;  provided we do not take too large a hierarchy, this factor is not necessarily very large. Furthermore, for the electroweak precision observables $S$ and $T$, we expect the effect of the running is either higher-loop or power-suppressed in $m_\chi/M_F$ and therefore a subdominant modification to the coefficients generated at the high-scale. This is in contrast to, e.g., the case considered in \citeR{Henning:2014gca}, where the operators responsible for $S$ and $T$ are not generated at the high-scale, and arise purely from running: in that case, it is important to find the contributions from running to the corresponding Wilson coefficients because $S$ and $T$ constraints are so strong.  Our neglect of the RGE in this model is also numerically justified, at least for the $S$ and $T$ parameters, as we will show in comparing EFT and full-theory one-loop results.
 
We thus move on to the matching at the low-scale $\mu = m_\chi$ where we integrate the $\chi$ out, working consistently correct to one-loop in the matching. The operators ${\cal O}_i$ for $i \in \{ 5,\, 6A,\, 6B,\, 7A,$ etc.\} are thereby removed from the Lagrangian and the Wilson coefficients $B_{i}$ for the dimension-6 operators with only SM field content receive new contributions in the EFT valid below $\mu = m_\chi$. We call these shifted Wilson coefficients in the low-scale EFT $C_{i}$; they are shown in \tabref{low_scale_coefficients} \emph{assuming no running or operator mixing} between the scales $\mu = M_F$ and $\mu = m_\chi$.
 
While the basis of operators shown in \tabref{high_scale_operators} was convenient for the matching computation, it still contains the various redundancies we mentioned earlier.  It also does not happen to be the most convenient for the computation of the $S$ and $T$ parameters, and $\sigma_{Zh}$. The $S$ and $T$ parameters are each simply related to the Wilson coefficient of a single operator when these redundancies are removed. In addition, \citeR{Craig:2014una} contains a computation of $\sigma_{Zh}$ in terms of a different basis of operators, and it useful to transform to this basis to enable us to perform a cross-check on our own independent computation of $\sigma_{Zh}$.%

As ${\cal O}_{(2,4)}$ is non-standard in the literature, we eliminate it via the identity 
\begin{equation}
	{\cal O}_{(2,4)} = 2 | D^2 H |^2 - {\cal O}_{(BB)} - 2 {\cal O}_{(BW)} - {\cal O}_{(WW)} - 2{\cal O}_{(B)} - 2{\cal O}_{(W)},
\label{eq:O24}
\end{equation}
plus irrelevant total-derivative terms. We are also free to make use of the SM EOM for $H$ to trade out $|D^2 H|^2$ for corrections to the coefficients of the operators $|H|^2, |H|^4, |H|^6$ and the SM Yukawa Lagrangian ${\cal L}_{\textsc{sm}}^{\textsc{Yukawa}}$, in addition to two new types of dimension-6 operators---$H^\dagger H \times {\cal L}_{\textsc{sm}}^{\textsc{Yukawa}}$ and a number of four-fermion contact operators proportional to two powers of the Yukawa couplings---as well as dimension-8 corrections, the latter of which we consistently neglect. The corrections to the $|H|^2, |H|^4$ and ${\cal L}_{\textsc{sm}}^{\textsc{Yukawa}}$ terms are absorbed into unobservable shifts of the SM parameters $\mu^2,\lambda$ and $\Gamma_i$ (the SM Yukawa coupling matrices), but the remaining additional terms here lead in principle to observable effects.

Our last step is to rescale the $H$ field to obtain a canonically normalized covariant derivative term (i.e., we absorb the coefficient of $|D_\mu H|^2$, $(1 + C_{(2,2)})$, into the $H$ field) and re-define the SM couplings [$\mu^2$, $\lambda$, and $\Gamma_i$] to absorb both the shifts proportional to $C_{(2,2)}$, as well as the shifts to the SM parameters arising from $C_{(4,0)}$ and $C_{(2,0)}$. We note that in the coefficients of all dimension-6 operators, the rescaling of the $H$ field can simply be ignored as the leading coefficients are already linear in the $C_i$: any correction resulting from the $H$ rescaling is therefore proportional to $C_{(2,2)} \times C_i$ and thus formally counted as a two-loop effect, since a single $C_i$ insertion is formally counted as a one-loop correction. In other words, the coefficients $C_{(2,2)}$, $C_{(4,0)}$ and $C_{(2,0)}$ can simply be ignored in an analysis correct to one-loop (this assumes a degree of tuning in the bare value of $\mu^2$ such that we still obtain the correct measured Higgs mass).

\begingroup
\renewcommand{\arraystretch}{1.2}
\begin{table*}[t]
\centering
\caption{\label{tab:low_scale_coefficients} Wilson coefficients of the operators appearing in \tabref{high_scale_operators}, evaluated at $\mu = m_\chi$, the ``low scale,'' after the $\chi$ has been integrated out. These expressions are correct under the assumption of no operator running or mixing between the scales $\mu = M_F$ and $\mu = m_\chi$. Our sign conventions are as detailed in \tabref{high_scale_coefficients}. These expressions assume a hierarchy of scales $M_F \gg m_\chi$.}
\resizebox{\textwidth}{!}{
\begin{tabular}{rl}
\textbf{Coefficient} 	&	\textbf{Value} 	\\ \hline \hline
$C_{(2,0)}$ 		&$ - ( \kappa ^2  / 4 \pi ^2 )  \lb[ M_F^2 + M_F\, m_{\chi }  + m_{\chi}^2 \rb] \quad \text{ ( this appears to be exact ) } $ \\
$C_{(2,2)}$ 		&$+( \kappa ^2 / 16 \pi ^2 ) \big[ 1 - 2 \mcmf{} - 4 \mcmf{2} - 6 \mcmf{3} -8\mcmf{4}$\\
				&\phantom{$+( \kappa ^2 / 16 \pi ^2 ) \big[ 1$}\;$- 10\mcmf{5} - 12\mcmf{6} + \cdots \big] $ \\
$C_{(4,0)}$ 		&$ - ( \kappa^4 / 8\pi^2 ) \ \, \lb[ 1 + 4\mcmf{}  + 8 \mcmf{2} + 12 \mcmf{3}  +16 \mcmf{4} +\cdots \rb] $ \\
$C_{(4,2),A}$ 		&$ - ( \kappa^4 / 24\pi^2 ) \lb[ 7 + 8 \mcmf{} - 3 \mcmf{2} +\cdots \rb]$ \\
$C_{(4,2),B}$ 		&$ - ( \kappa^4 / 48\pi^2 ) \lb[ 5 - 2  \mcmf{} - 15 \mcmf{2} + \cdots \rb]$ \\
$C_{(4,2),C}$ 		&$ - ( \kappa^4 / 48\pi^2 ) \lb[ 1 - 22 \mcmf{} - 111 \mcmf{2} + \cdots \rb] $ \\
$C_{(6,0)}$ 		&$ - ( \kappa^6 / 12\pi^2 ) \lb[ 1 + 14 \mcmf{} + 51 \mcmf{2} + \cdots \rb] $ \\
$C_{(WW)}$ 		&$ - ( \kappa^2 / 144\pi^2 ) \lb[ 1 - 4\mcmf{} + 3\mcmf{2} + 10 \mcmf{3} + 41\mcmf{4} + \cdots \rb] $ \\
$C_{(BB)}$ 		&$ - ( \kappa^2 / 144\pi^2 ) \lb[ 1 - 4\mcmf{} + 3\mcmf{2} + 10 \mcmf{3} + 41\mcmf{4} + \cdots \rb] $ \\
$C_{(BW)}$		&$ - ( \kappa^2 / 72\pi^2 ) \lb[ 1 - 4\mcmf{} + 3\mcmf{2} + 10 \mcmf{3} + 41\mcmf{4} + \cdots \rb] $ \\
$C_{(2,4)}	$		&$ + ( \kappa^2 / 48\pi^2 ) \lb[ 1 - \mcmf{} - 3\mcmf{2} - 14 \mcmf{3} - 25 \mcmf{4} + \cdots \rb] $\\
$C_{(DB)}	$		&$ + ( 1 / 240\pi^2 ) $ \\
$C_{(DW)}$		&$ + ( 1 / 240\pi^2 ) $ \\
$C_{(W)}$ 		& $ + ( \kappa^2 / 72\pi^2 ) \lb[ 7 - 4 \mcmf{} - 12 \mcmf{2} + \cdots  \rb] $ \\
$C_{(B)}$ 			& $ + ( \kappa^2 / 72\pi^2 ) \lb[ 7 - 4 \mcmf{} - 12 \mcmf{2} +  \cdots  \rb] $ \\
$C_{(WWW)}$ 		&  Operator generated, but coefficient not computed. \\
				& \quad Irrelevant to the observables of interest.
\end{tabular}}
\end{table*}
\endgroup

\subsubsection{Precision electroweak observables}
\label{sect:singletdoubletEWPO}

Having eliminated ${\cal O}_{(2,4)}$ using \eqref{O24}, the $S$ and $T$ parameters can be read off from the new coefficients of the operators ${\cal O}_{(BW)}$ and ${\cal O}_{(4,2),B}$, respectively.\footnote{Again, we do not run the operator coefficients between $m_\chi$ and the $Z$-pole.} These operators modify the gauge boson vacuum polarization amplitudes $\Pi_{VV}^{\mu \nu}(q^2) = i g^{\mu \nu} \Pi_{VV}(q^2) + ...$ . Correct to linear order in the Wilson coefficients $C_i$, we find that the new physics contributions to $S$ and $T$ are:\footnote{The operators ${\cal O}_{(DB),(DW)}$ contribute only terms $\sim p^4$ to the gauge-boson two-point functions and hence do not contribute to the $S$, and $T$ parameters (at least if $S$ is defined per \eqref{deriv_S}); instead, these operators give rise to the $\hat{Y}$ and $\hat{W}$ operators of \citeR{Barbieri:2004qk}. However, we find the corresponding limits are not competitive with those from $S$ and $T$ and do not pursue this further.}
\renewcommand{\arraystretch}{1.0}
\begin{align}
S &\equiv \frac{4c_W^2s_W^2}{\alpha_e}\lb[  \dfrac{ \Pi_{ZZ}(M_Z^2) - \Pi_{ZZ}(0) }{M_Z^2}
											 - \dfrac{c_W^2-s_W^2}{c_Ws_W} \dfrac{ \Pi_{Z\gamma}(M_Z^2)}{M_Z^2}
											  - \dfrac{ \Pi_{\gamma\gamma}(M_Z^2) }{M_Z^2} \rb] \\
&\approx \frac{4c_W^2s_W^2}{\alpha_e}\lb[ \Pi_{ZZ}'(0) - \frac{c_W^2-s_W^2}{c_Ws_W} \Pi_{Z\gamma}'(0) - \Pi_{\gamma\gamma}'(0) \rb] \label{eq:deriv_S} \\
&= -\frac{4c_Ws_W}{\alpha_e} \Pi_{W^3B}'(0) \\
&\approx - 4\pi \frac{v^2}{M_F^2} \lb[ C_{(BW)} - 2 C_{(2,4)} \rb],  
\end{align}
and
\begin{align}
T&= \frac {\rho-1}{\alpha_e}\\
&\equiv \frac{1}{ \alpha_e M_W^2} \bigg[ \Pi_{W^+W^-}(0) - \Pi_{W^3W^3}(0) \bigg] \\[1ex]
& = \frac{1}{\alpha_e M_W^2} \bigg[ \Pi_{W^+W^-}(0) - c_W^2 \Pi_{ZZ}(0) \bigg] \label{eq:2} \\[1ex]
&\approx -\frac{1}{2\alpha_e} \frac{v^2}{M_F^2} C_{(4,2),B},
\end{align}
where in arriving at Eqs.~(\ref{eq:deriv_S}) and (\ref{eq:2}), we have made use of $\Pi_{Z\gamma}(0) = \Pi_{\gamma\gamma}(0) = 0$.  The $U$ parameter is identically zero at dimension 6; the first non-zero contribution occurs at dimension 8.

Substituting the explicit forms of the Wilson coefficients, we have
\renewcommand{\mcmf}[1]{\dfrac{ m_\chi^{#1} }{ M_F^{#1} } }
\begin{align}
S 	&\approx\phantom{\frac{\alpha_\kappa}{\alpha_e}\, } \frac{ 2\kappa^2}{9\pi} \frac{v^2}{M_F^2} \lb[  1- \dfrac{7}{4} \mcmf{} - \dfrac{3}{2} \mcmf{2} - 8 \mcmf{3} - \dfrac{17}{2} \mcmf{4} + \cdots  \rb]  \\[2ex]
T 	&\approx \frac{\alpha_\kappa}{\alpha_e} \, \frac{5 \kappa^2 }{24\pi }   \frac{v^2}{M_F^2} \lb[ 1 - \frac{2}{5}  \frac{m_\chi}{M_F}- 3 \frac{m_\chi^2}{M_F^2} + \cdots \rb] \quad \text{where} \quad \alpha_\kappa \equiv \frac{\kappa^2}{4\pi} \\[2ex]
U &= 0 + \text{(dim-8)}
\end{align}
The parametric enhancement (or suppression, depending on the relative sizes of $\alpha_\kappa$ and $\alpha_e$) of $T$ compared to $S$ can be understood as follows. For $M_F\gg m_\chi$, $\Pi_{W^+W^-}(0) - c_W^2 \Pi_{ZZ}(0) \sim g^2 (\kappa^2  v^2) ( \kappa^2 v^2 / M_F^2)$, while $\Pi'_{ZZ}(0) \sim (g^2+(g')^2) ( \kappa^2v^2 / M_F^2)$. This is determined by the mass dimension of the quantity concerned and the fact that, in our model, a diagrammatic new-physics contribution proportional to the Higgs vev $v$ must always actually be proportional to $\kappa v$ owing to the structure of the $H$-$F$-$\chi$ coupling. Therefore, $T \sim ( g^2 v^2 / M_W^2 ) (\kappa^2 /\alpha_e) ( \kappa^2 v^2 / M_F^2) \sim ( \alpha_\kappa / \alpha_e)  ( \kappa^2 v^2 / M_F^2 )$ while $S \sim (c_W^2s_W^2/ \alpha_e ) \Pi'_{ZZ}(0) \sim \kappa^2v^2/ M_F^2$ since $\alpha_e/c_W^2s_W^2 \sim g^2 + (g')^2$. The parametric dependence on $\kappa$ can also be read off from the form of the dimension-6 operators giving rise to $S$ and $T$. Upon EWSB, the $S$ parameter arises from ${\cal O}_{(BW)} = H^\dagger \hat B_{\mu \nu} \hat W^{\mu \nu} H$. As a result, diagrammatic contributions to $S$ come with two powers of the Higgs vev $v$ and hence two powers of $\kappa v$, since the two derivatives in the field strengths are removed in the definition of $S$ as a momentum derivative of the two-point function. The $T$ parameter arises, upon EWSB, from ${\cal O}_{(4,2),B} = |H^\dagger D_\mu H|^2$; the diagrammatic contributions to $T$ thus come with four powers of the Higgs vev $v$ and hence four powers of $\kappa v$ (two of the Higgs vev factors are later cancelled by the $M_W^{-2}$ in the definition of $T$).  Therefore, $T \sim \kappa^4$ and $S\sim \kappa^2$.

\begin{figure*}[t]
\centering
\includegraphics[width = 0.49 \textwidth]{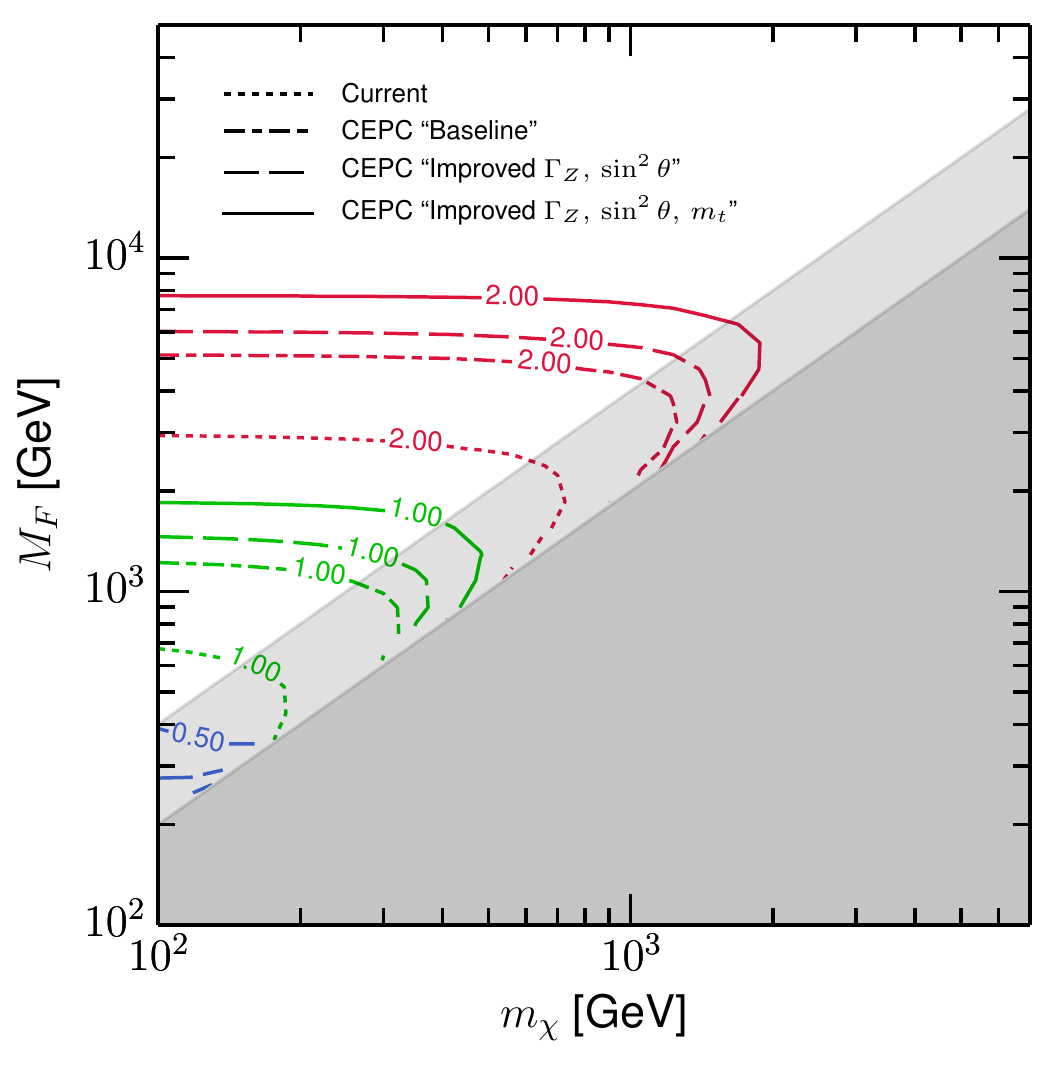}
\includegraphics[width = 0.49 \textwidth]{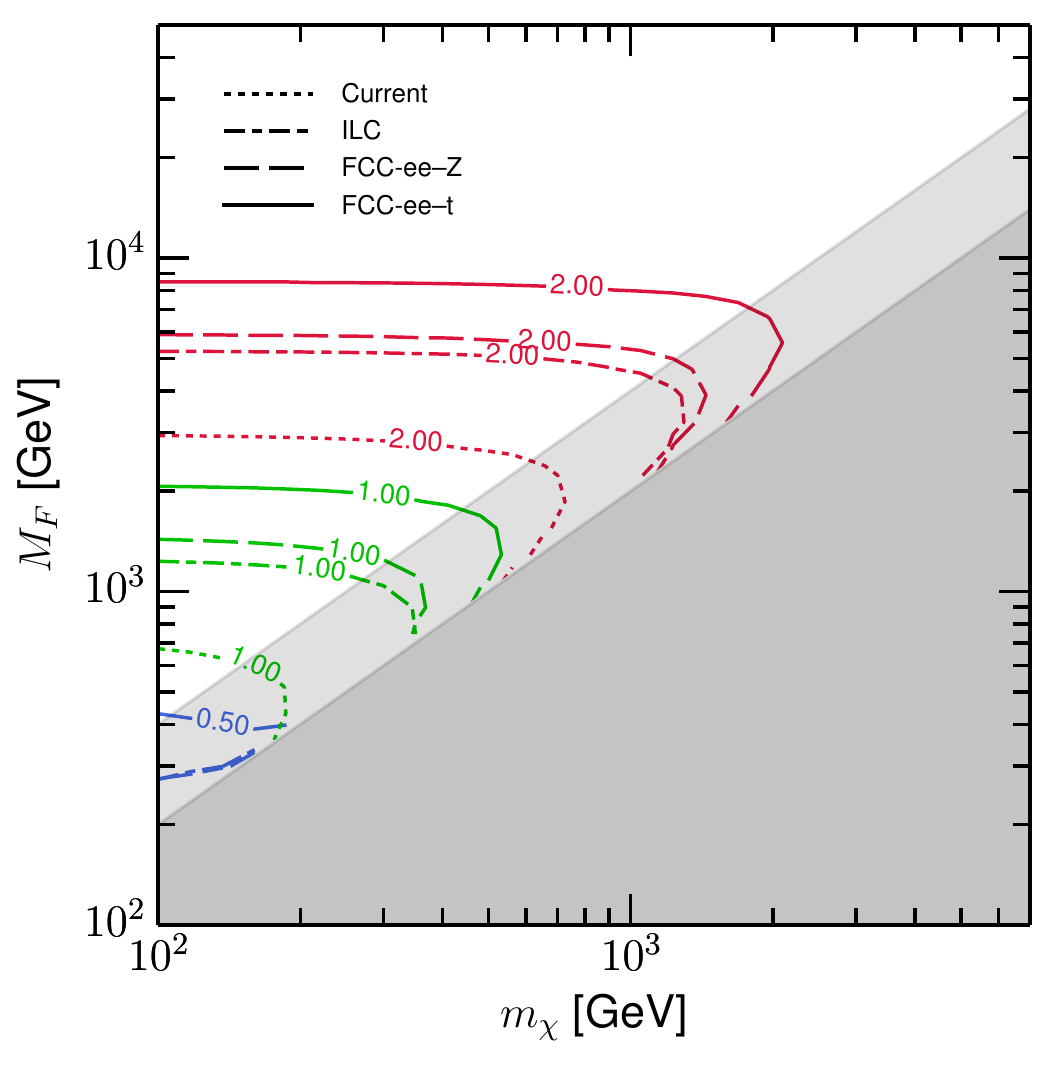}
\caption{ \footnotesize \label{fig:ewpo_EFT} Two-parameter 95\% confidence exclusion regions ($-2 \log[ {\cal L}/{\cal L}_0] \gtrsim 5.99$) from measurement of the precision electroweak variables $(S,T)$, using results computed in the EFT.  We present these as boundaries in the allowed mass parameter space for fixed representative values of $\kappa$, as annotated on each line; the \emph{unshaded} region to the lower-left of these lines is excluded for the given value of $\kappa$. To compare EFT results with the full-theory calculation, see \figref{ewpo_full}. The light shaded region, $M_F/4 \lesssim m_\chi \lesssim M_F/2$, denotes the region where the EFT begins to break down: the error in the EFT result for $T$ compared to the $v^2/M_F^2$ piece of the full result (i.e., the ``dimension-6 part of'' the full result) is $\sim 20\%$ at $m_\chi \sim  M_F/4$, reaches $\sim50\%$ at $m_\chi \sim M_F/3$, and becomes $>100\%$ before $m_\chi \sim M_F/2$. In the dark-shaded region, $m_\chi \gtrsim M_F/2$, the results have consequently been masked as they are invalid. The various line styles correspond to the current constraints and various projected constraints on $(S,T)$ for the proposed CEPC collider (left plot) and the proposed ILC and FCC-ee colliders (right plot); see \tabref{sensitivities}.}
\end{figure*}

Note that an indirect effect of the dimension-6 operators may be to imply shifts to the Lagrangian coupling constants (e.g., the gauge couplings) away from their SM reference values, as we will discuss when we turn to the Higgsstrahlung process in the following section. However, since the leading results for $S$ and $T$ are already proportional to $C_i$ and hence formally one-loop, any coupling shift can be dropped as a higher-order correction ($S=T=0$ in the SM at tree-level). Moreover, the coupling constant shifts will imply a shift in the numerical values of the one-loop computations of the SM contributions to the $S$ and $T$ parameters, but these are again formally at least two-loop effects. Since we work to one-loop accuracy, we neglect these contributions and the value of $\alpha_e$ appearing in $T$ is taken to be its SM value (we take it at the $Z$ pole).

In \figref{ewpo_EFT}, we show the 95$\%$ confidence exclusion regions in the mass parameter space from measurements of the electroweak precision variables $(S,T)$, for fixed representative values of $\kappa$. These results are computed in the EFT, and are shown for both current limits (LEP+SLD) and for the proposed CEPC, ILC and FCC-ee colliders with the sensitivities given in \tabref{sensitivities}. 

In the region of parameter space under consideration in \figref{ewpo_EFT}, the limits are driven almost exclusively by the size of the $T$ parameter, owing to its parametric enhancement $\sim \alpha_\kappa / \alpha_e$ over the $S$ parameter. In the regime where the EFT is valid, $m_\chi \lesssim \frac{1}{4} M_F$, the exclusion reach is also largely insensitive to the value of $m_\chi$, as one might expect. Already with current (LEP+SLD) constraints on $S,T$, masses $M_F$ below about 675~GeV can be ruled out for $\kappa \sim 1$; due to the $\kappa^4$ dependence of $T$, this lower limit increases to around 2.9~TeV for a coupling $\kappa \sim2$, but there is essentially no limit in the regime where the EFT is valid if $\kappa \sim 0.5$. 

With even the ``baseline'' proposal for the CEPC, these lower limits increase by a factor of approximately 1.75 owing to the factor $\sim 3.5$ times stronger limits on $S$ and $T$ in this scenario as compared to the current bounds \cite{Fan:2014vta}.  For the CEPC scenario with improved measurements of $\Gamma_Z$ and $\sin^2\theta$ only, the lower limits on $M_F$ increase by a factor of $\sim 2$ compared to the present limits, rising to $\sim 1.5$~TeV (6~TeV) for $\kappa \sim 1.0$ (2.0), in the limit where $m_\chi \ll M_F$. This results from the tightening in the limits on $T$ (and $S$) by factor of $\sim 4$ compared to the present constraints \cite{Fan:2014vta}.
For the best-case scenario for the CEPC, with improved measurements of $\Gamma_Z$, $\sin^2\theta$, and $m_t$, the lower limits on $M_F$ increase by a factor of $\sim 2.7$ compared to the present limits (i.e., about 30\% over the previous scenario), rising to $\sim 1.8$~TeV (7.7~TeV) for $\kappa \sim 1.0$ (2.0), in the limit where $m_\chi \ll M_F$. This results from the tightening in the limit on $T$ by factor of $\sim 7$, and the limit in $S$ by a factor of $\sim 5$ compared to the present limits (i.e., the $m_t$ measurement roughly halves again the measurement uncertainty on $T$ compared to the previous scenario, and also improves the $S$ constraint) \cite{Fan:2014vta}.

The projected sensitivities of the ILC and FCC-ee show improvements in the lower limits on $M_F$ which are broadly similar to the various CEPC scenarios. The ILC projection and CECP ``baseline'' projection are very similar, as are the FCC-ee-Z limits and the CEPC limits with improved measurements of $\Gamma_Z$ and $\sin^2\theta$ only. Finally, the FCC-ee-t lower-limits on $M_F$ (for $\kappa \sim 1.0$ and $\kappa \sim 2.0$ they are $2.1$~TeV and $8.5$~TeV, respectively) are slightly stronger than the best-case CEPC scenario, but are nevertheless still broadly similar.

\subsubsection{Higgsstrahlung}
\label{sect:singletdoubletHiggs}
The computation of shifts to the Higgsstrahlung cross-section $\sigma_{Zh}$ is more involved than that required to obtain the $S$ and $T$ parameters. There are a large number of operators which directly contribute to a shift in the cross-section, both by way of contributions to  wavefunction renormalization of the $h,Z$ fields in the $hZ^\mu Z_\mu$ coupling, and the introduction of additional diagrams. Furthermore, $\sigma_{Zh}$ is nonzero at tree-level in the SM and so the one-loop new-physics shifts to the relationships between SM input parameters and SM Lagrangian parameter values must be accounted for here to maintain a result consistently correct to one-loop order: the standard set of SM input parameters for high-precision work are $(G_F, m_Z, \alpha_e)$ and the dimension-6 operators impact the processes (e.g., muon decay at $q^2=0$) used to relate the numerical values of these parameters to the values of Lagrangian parameters, leading to shifts in the Lagrangian coupling constants away from their SM reference values~\cite{Burgess:1993vc}.

The requisite computation of $\sigma_{Zh}$ has actually recently appeared in the literature in \citeR{Craig:2014una}. We have nevertheless independently repeated the computation of $\sigma_{Zh}$ in the EFT as a cross-check; we find complete agreement with their results. In order for us to make reference to the results of \citeR{Craig:2014una}, we transform to the same basis of operators used there. In addition to the elimination of ${\cal O}_{(2,4)}$ described above,
four further sets of manipulations on our basis of operators are required: (a) we re-write the operator 
\begin{align}
{\cal O}_{(4,2),B}=| H^\dagger D_\mu H |^2 = - \frac{1}{4} ( H^\dagger \lrD{\mu} H )^2 + \frac{1}{4} ( \partial_\mu |H|^2 )^2;
\end{align}
(b) we rewrite the operators ${\cal O}_{(DB)}$ and  ${\cal O}_{(DW)}$, up to total derivative terms, as
\begin{align}
{\cal O}_{(DB)} &= -\frac{(g')^2}{2} ( \partial_\mu B_{\nu\rho} ) ( \partial^\mu B^{\nu\rho} ) = -(g')^2 ( \partial_\mu B^{\mu\nu} )^2, \quad \text{and} \\
{\cal O}_{(DW)} &= \Tr{ [ D_\mu, \hat{W}_{\nu\rho} ] [ D^\mu, \hat{W}^{\nu\rho} ] } = 2\, \Tr{ [ D_\mu , \hat{W}^{\mu\rho} ] [ D_\nu , {\hat{W}^{\nu}}_{\ \; \rho\, } ] } + 4\, \Tr{{\hat{W}_{\rho}}^{\ \; \nu \,} {\hat{W}_{\nu}}^{\ \; \mu\, } {\hat{W}_{\mu}}^{\ \; \rho \, }   }, \label{eq:ODWremoval}
\end{align}
and utilise the SM EOM for $\hat{B}_{\mu\nu}$ and $\hat{W}_{\mu\nu}$ to rewrite the first term on the RHS of each of these relationships; and (c) we make two manipulations to remove ${\cal O}_{(4,2),C}$: 
\begin{align}
H^\dagger H | D_\mu H |^2  = - \frac{1}{2} ( \partial_\mu |H|^2 )^2 - \frac{1}{2} H^\dagger H \lb[ H^\dagger ( D^2 H ) + ( D^2 H)^\dagger H \rb],
\end{align}
followed by the substitution of the SM EOM to rewrite $D^2H$. Finally, (d) we rewrite the operators ${\cal O}_{(B)}$ and ${\cal O}_{(W)}$, up to total-derivative terms, as
\begin{align}
{\cal O}_{(B)}& = \frac{1}{2} {\cal O}_{(BB)} + \frac{1}{2}{\cal O}_{(BW)} - \frac{g'}{4} \lb[ H^\dagger i \lrD{\nu}\, H \rb] \lb( \partial_\mu B^{\mu\nu} \rb), \quad \text{and} \\
{\cal O}_{(W)}& = \frac{1}{2} {\cal O}_{(WW)} + \frac{1}{2}{\cal O}_{(BW)} - \frac{g}{4} \lb[ H^\dagger i \sigma^a \lrD{\nu}\, H \rb] \lb( D_\mu W^{a,\, \mu\nu} \rb),
\end{align}
and then utilise the SM EOM for $\hat{B}_{\mu\nu}$ and $\hat{W}_{\mu\nu}$ to re-write the last term on the RHS in each line, and the SM EOM for $H$ to re-write some factors of $D^2H$ which appear upon doing so. This is followed by a final re-definition of the SM Higgs-Lagrangian parameters, to absorb some unobservable shifts to the SM parameters which occur during this process.

When the dust settles, we are left with 
\begin{align}
{\cal L} &= {\cal L}_{\textsc{sm}} + \sum_j \frac{ C_j }{M_F^2} {\cal O}^{(6)}_{j}, \label{eq:shifted_Lag}
\end{align}
where the dimension-6 operators ${\cal O}^{(6)}_{j}$ and Wilson coefficients $C_j$, correct to linear order in the original $C_{(\,\cdots\!\;)}$, are listed in \tabref{shifted_Lag_ops}, and ${\cal L}_{\textsc{sm}}$ takes the same form as the SM Lagrangian, except that all the parameters are now understood to be defined so as to absorb any of the unobservable shifts which occurred as a result of the manipulations just described.\footnote{These unobservable shifts can consistently be ignored; they are not the same as the physically relevant shifts which need to be accounted for in relating the input parameters to the values of the Lagrangian parameters.} As far as the $e^+e^- \rightarrow Zh$ process is concerned, all of the operators in \tabref{shifted_Lag_ops} listed above the horizontal line contribute, either through a shift in the SM input parameters or through the addition of a new term in the amplitude itself.\footnote{Note that $C_{(WW)} = C_{(BB)} = \frac{1}{2}C_{(BW)}$ and $C_{(W)} = C_{(B)}$ in our model, so the Wilson coefficients of the operators $H^\dagger \hat{W}_{\mu\nu} \hat{W}^{\mu\nu} H$ and $H^\dagger \hat{B}_{\mu\nu} \hat{B}^{\mu\nu} H$ remain equal, and equal to one-half of the Wilson coefficient of the operator $H^\dagger \hat{B}_{\mu\nu} \hat{W}^{\mu\nu} H$. This ultimately follows from the fact that there are no new Higgs-charged-charged vertices in the full theory picture, and guarantees that there is no leading-order (one-loop) correction to the $h\rightarrow\gamma\gamma$ decay rate. This would not be true if we had introduced a charged partner to the $\chi$ to maintain custodial symmetry, as we discussed earlier.}

\begingroup
\renewcommand{\arraystretch}{1.2}
\begin{table*}[t]
\centering
\caption{ \label{tab:shifted_Lag_ops} Operators ${\cal O}_j^{(6)}$ and corresponding Wilson coefficients $C_j$ appearing in the effective Lagrangian which we utilise to compute $\sigma_{Zh}$ [see \eqref{shifted_Lag}]. Only the operators listed above the horizontal line contribute a shift at leading order to $e^+e^- \rightarrow Zh$, either through a shift in the relationships between the SM input parameters and the Lagrangian parameters, or through the addition of a new term in the amplitude itself. $\lambda, \mu$ and the Yukawa matrices $\Gamma_i$ are understood to take their unobservably shifted values. We define $\hat{W}_{\mu\nu} \equiv i g W^a_{\mu\nu} t^a$, $\hat{B}_{\mu\nu} \equiv i g' B_{\mu\nu} Y$, $\lrD{\mu} \; \; \equiv D_{\mu} \, - \stackrel{\leftarrow}{D}_{\mu}$ and $\sigma^a \lrD{\mu} \; \;  \equiv \sigma^a D_{\mu} \, - \stackrel{\leftarrow}{D}_{\mu}\! \sigma^a$. Superscript $p,q$ are generation indices; $i,j$ are $SU(2)$-fundamental indices; and $a$ is an $SU(2)$-adjoint index---all are summed over if repeated. } 
\resizebox{\textwidth}{!}{
\begin{tabular}{lll}
\textbf{Operator} $\bm{{\cal O}_j^{(6)}}$									& \textbf{Wilson coefficient}  $\bm{C_j}$									\\ \hline \hline
$\tfrac{1}{2} ( \partial_\mu |H|^2 )^2$ 							& $-C_{(4,2),C} + \frac{1}{2} C_{(4,2),B} + C_{(4,2),A} - \frac{3}{4}g^2 C_{(W)}$\\
														&\qquad $ + \frac{3}{2}g^4 C_{(DW)} + \frac{3}{2} g^2 C_{(2,4)}$	\\[1ex]
$\frac{1}{4} ( H^\dagger \lrD{\mu} H )^2$ 						& $-C_{(4,2),B} - \frac{1}{2}(g')^2  C_{(B)}  + (g')^4 C_{(DB)} +(g')^2 C_{(2,4)} $										\\[1ex]
$H^\dagger \hat{W}_{\mu\nu} \hat{W}^{\mu\nu} H$ 				& $C_{(WW)} + \frac{1}{2} C_{(W)} - 2C_{(2,4)} $												\\[1ex]
$H^\dagger \hat{B}_{\mu\nu} \hat{B}^{\mu\nu} H$ 					& $C_{(BB)} + \frac{1}{2} C_{(B)}- 2C_{(2,4)} $												\\[1ex]
$H^\dagger \hat{B}_{\mu\nu} \hat{W}^{\mu\nu} H$					& $C_{(BW)} +\frac{1}{2} \lb( C_{(W)} + C_{(B)} \rb) - 4 C_{(2,4)}$ 											\\[1ex]
$( H^\dagger i\lrD{\mu} H ) \lb(\sum_{f_L,f_R} Y_f \bar{f} \gamma^\mu f\rb)$ 					& $-(g')^4 C_{(DB)} + \frac{1}{4}(g')^2 \lb( C_{(B)} - 2 C_{(2,4)} \rb)$				\\[1ex]
$( H^\dagger \sigma^a i \lrD{\mu} H ) \lb( \bar{L}_L^p \gamma^\mu \sigma^a L_L^p \rb)$ 		& $-\frac{1}{2} g^4C_{(DW)} + \frac{1}{8} g^2 \lb( C_{(W)} - 2 C_{(2,4)} \rb)$		\\[1ex]
$\lb( \bar{L}_L^p \gamma_\mu \sigma^a L_L^p \rb)^2$									& $-\frac{g^4}{4} C_{(DW)}$		\\[1ex] \hline
$|H|^6$													& $+ C_{(6,0)}	+  \lambda \Big( 8 C_{(2,4)} +2 C_{(4,2),C} -2 g^4 C_{(DW)} $	\\
														&\phantom{$+ C_{(6,0)}	+  \lambda \Big($}\quad $+g^2 C_{(W)} - 2g^2  C_{(2,4)}  \Big)$			\\[1ex]
$\Tr{\hat{W}_{\mu\nu} \hat{W}^{\nu\sigma} {\hat{W}_{\sigma}}^{\;\; \mu} }$  		& $ C_{(WWW)} + 4 C_{(DW)}$								\\[1ex]
$( H^\dagger H ) \lb[ \begin{array}{l} \lb(\bar{L}^j\, \Gamma_e\, e\rb) H^j + \lb( \bar{Q}^j\, \Gamma_d\, d \rb) H^j \\ + \lb( \bar{Q}^j\, \Gamma_u\,   u \rb) \epsilon_{jk} \lb(H^\dagger \rb)^k  \end{array}\rb] + \text{h.c.}$ & 	$ \frac{1}{2} C_{(4,2),C} -\frac{1}{2} g^4 C_{(DW)} + \frac{1}{4} g^2C_{(W)} + \frac{1}{2}\lb( 8 \lambda - g^2\rb)C_{(2,4)} $ \\[4ex]
$( H^\dagger \sigma^a i \lrD{\mu} H ) \lb( \bar{Q}_L^p \gamma^\mu \sigma^a Q_L^p \rb)$		& $-\frac{1}{2}g^4 C_{(DW)}+ \frac{1}{8} g^2 \lb( C_{(W)} - 2 C_{(2,4)} \rb)$		\\[1ex]
$\lb(\sum_{f_L,f_R} Y_f \bar{f} \gamma_\mu f\rb)^2$ 									& $-(g')^4 C_{(DB)}$				\\[1ex]
$\lb( \bar{Q}_L^p \gamma_\mu \sigma^a Q_L^p \rb)^2$									& $-\frac{1}{4} g^4C_{(DW)}$		\\[1ex]
$\lb( \bar{L}_L^p \gamma_\mu \sigma^a L_L^p \rb)\lb( \bar{Q}_L^q \gamma^\mu \sigma^a Q_L^q \rb)$  & $-\frac{1}{2}g^4 C_{(DW)}$		\\[1ex]
$\lb[ \begin{array}{l} 
					 		( \bar{L} \Gamma_e e )( \bar{e} {\Gamma_e}^\dagger L ) + ( \bar{Q} \Gamma_d d)(\bar{d} {\Gamma_d}^\dagger Q) \\
							+ (\bar{Q}\Gamma_u u )(\bar{u} {\Gamma_u}^\dagger Q ) \\+ ( (\bar{L}\Gamma_e e)(\bar{d} {\Gamma_d}^\dagger Q) + \text{h.c.} ) \\
							- ( (\bar{L}^j\Gamma_e e) \epsilon_{jk} (\bar{Q}^k\Gamma_u u) + \text{h.c.} ) \\
							- ( (\bar{Q}^j\Gamma_d d) \epsilon_{jk} (\bar{Q}^k\Gamma_u u) + \text{h.c.} ) 
							\end{array} \rb]$ & $2C_{(2,4)}$
\end{tabular}}
\end{table*}
\endgroup

\begingroup
\renewcommand{\arraystretch}{1.2}
\begin{table}[t]
\centering
\caption{ \label{tab:dictionary} A dictionary to convert our Wilson cofficient results from \tabref{low_scale_coefficients} to the Wilson coefficients defined in \citeR{Craig:2014una}. } 
\resizebox{\textwidth}{!}{
\begin{tabular}{ll}
\textbf{Wilson Coefficient in \citeR{Craig:2014una}} 	&	\textbf{Value per our \tabref{low_scale_coefficients}}									\\ \hline
$c_{WW}$								& 	$-\frac{1}{8} \lb( 2C_{(WW)} + C_{(W)} -4 C_{(2,4)}	\rb) $			\\
$c_{BB}$									& 	$-\frac{1}{8} \lb(2 C_{(BB)} + C_{(B)} - 4C_{(2,4)}		\rb)$			\\
$c_{WB}$									& 	$-\frac{1}{8} \lb( 2C_{(BW)}  + C_{(W)} + C_{(B)} - 8C_{(2,4)}	\rb)$			\\
$c_H$									&	$-C_{(4,2),C} + \frac{1}{2} C_{(4,2),B} + C_{(4,2),A} $\\
										&	\qquad $+ \frac{3}{2}g^4 C_{(DW)} - \frac{3}{4}g^2\lb( C_{(W)} - 2C_{(2,4)} \rb)$	\\
$c_T$									&	$ -\frac{1}{2} C_{(4,2),B} + \frac{1}{2}(g')^4 C_{(DB)}- \frac{1}{4}(g')^2 \lb( C_{(B)} -2 C_{(2,4)} \rb)$		\\
$c_L^{(3)l}$								&	$-\frac{1}{2}g^4 C_{(DW)} +\frac{1}{8}g^2 \lb( C_{(W)} -2C_{(2,4)} \rb)$							\\
$c_{LL}^{(3)l}$								&	$-\frac{1}{4}g^4 C_{(DW)}$							\\
$c_L^{l}$									&	$\frac{1}{2}(g')^4 C_{(DB)} - \frac{1}{8}(g')^2 \lb( C_{(B)} - 2C_{(2,4)} \rb)$						\\
$c_R^{e}$									&	$(g')^4 C_{(DB)} - \frac{1}{4}(g')^2 \lb( C_{(B)} - 2C_{(2,4)} \rb)$								
\end{tabular}	}
\end{table}
\endgroup

The results of \citeR{Craig:2014una} can then be used to read off the value of $\sigma_{Zh}$; in \tabref{dictionary}, we supply an explicit dictionary to transform our Wilson coefficients into the notational conventions of \citeR{Craig:2014una}. In all our results, we utilize the $(G_F,M_Z,\alpha_e)$ input-parameter set.

Apart from the independent re-computation of $\sigma_{Zh}$ beginning from the basis of operators and Wilson coefficients in \tabref{shifted_Lag_ops} to cross-check against the results of \citeR{Craig:2014una}, we also verified the correctness of the fairly complicated manipulations which were used to remove the operator ${\cal O}_{(DW)}$ (see \eqref{ODWremoval} and the text following) in arriving at \eqref{shifted_Lag}. To do so, we augmented the SM Lagrangian with ${\cal O}_{(DW)}$ alone and considered its effects on $\sigma_{Zh}$ directly, rather than eliminating this operator by EOM.

Furthermore, since the ILC measurement projections assume a polarized electron beam with $(P_{e^-},P_{e^+}) = ( -0.8,+0.3)$ \cite{Asner:2013psa}, in the course of our independent computation, we also computed the polarised-beam cross-section. In the notation of \citeR{Craig:2014una}, the polarized cross-section is obtained from the unpolarized cross-section by making the following simple replacements in the quantities defined either in their eq.~(2.3) or in their table 2: multiply every appearance of $g_L^2$ in their $F_{\textsc{sm}}$ and $F_{1,2}$, as well as every appearance of $g_L$ in their $F_{3,4}$, by $(1-P_{e^-})(1+P_{e^+})$. Similarly, multiply every appearance of $g_R^2$ in their $F_{\textsc{sm}}$ and $F_{1,2}$, as well as every appearance of $g_R$ in their $F_{3,5}$, by $(1+P_{e^-})(1-P_{e^+})$. No changes are needed to the factors appearing explicitly in their eq.~(3.9). 

\begin{figure*}[t]
\centering
\includegraphics[width = 0.49\textwidth]{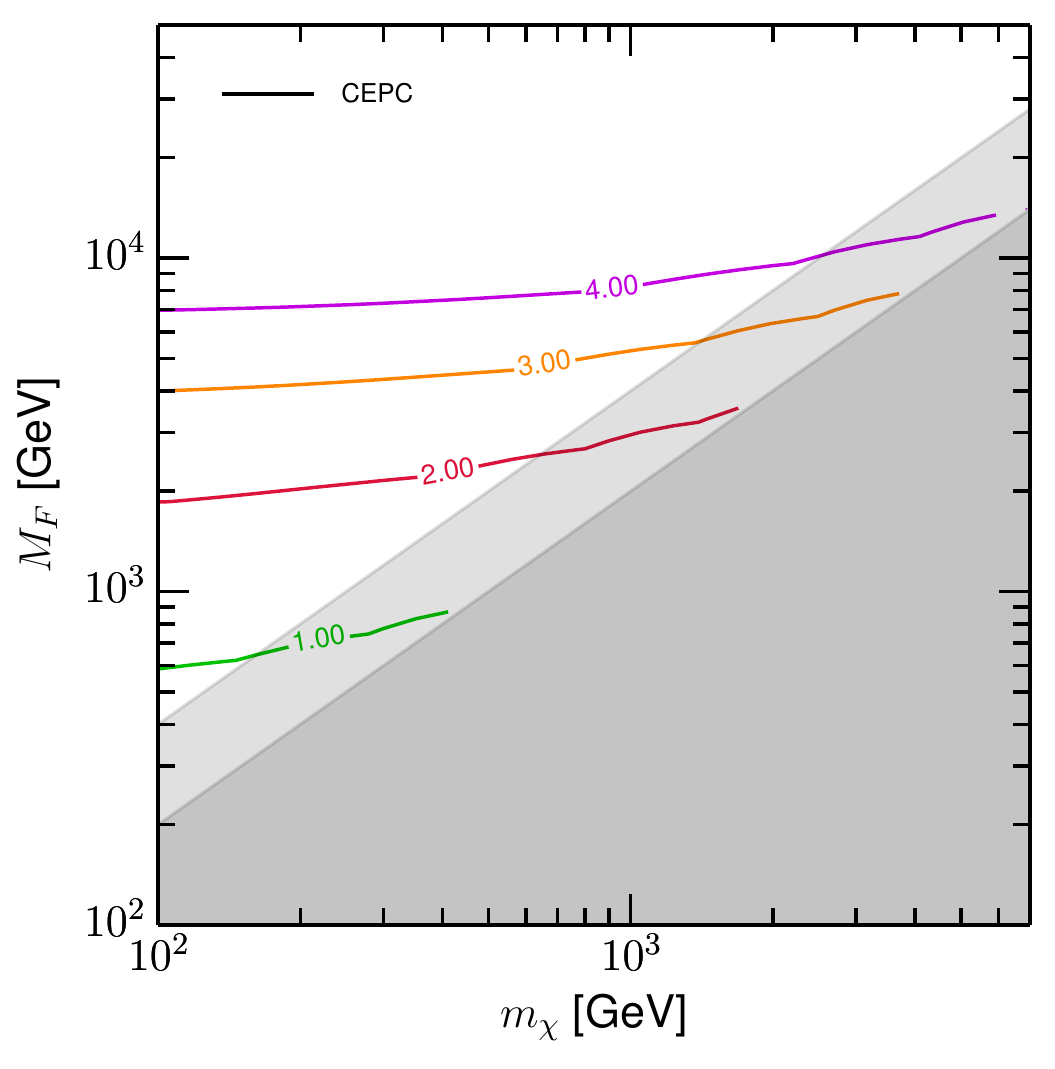}
\includegraphics[width = 0.49\textwidth]{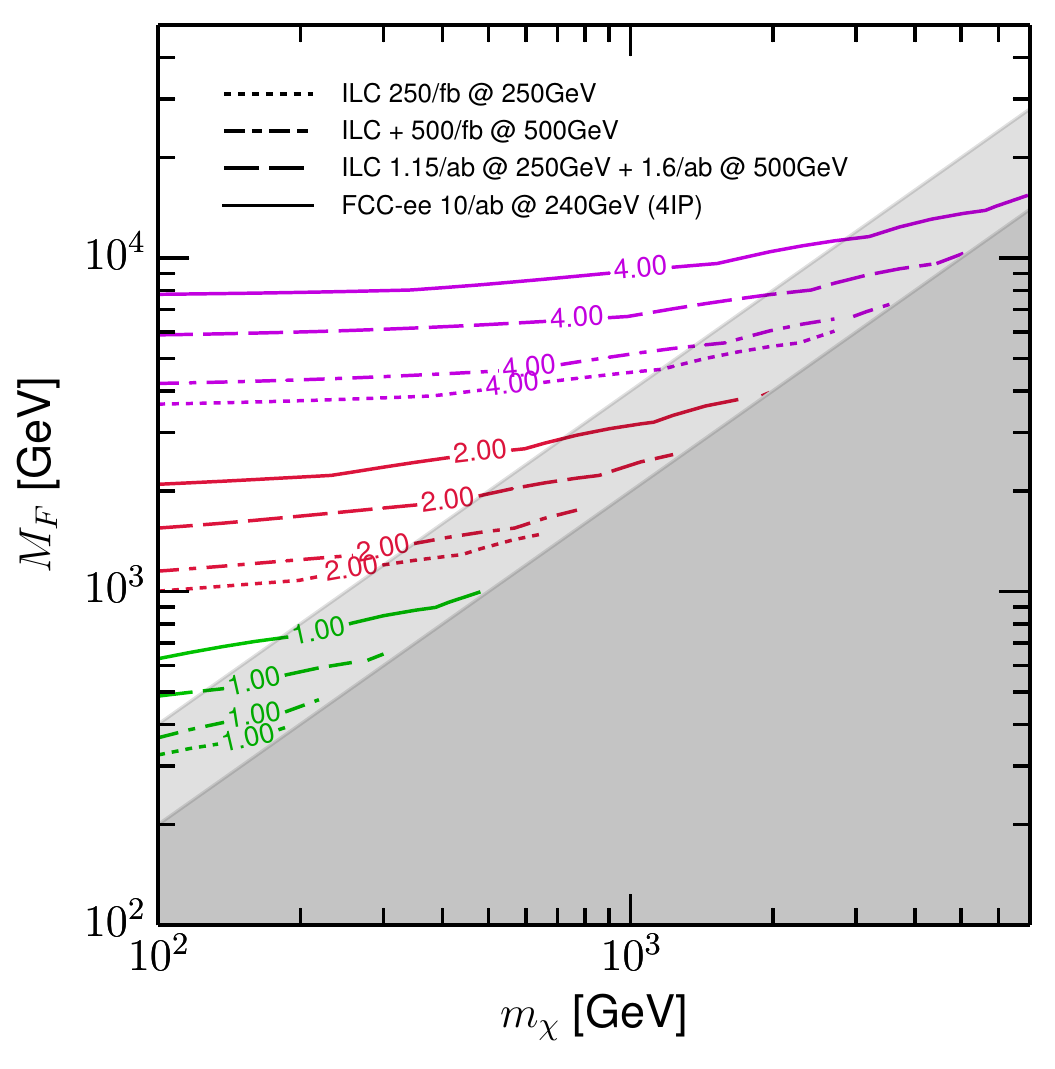}
\caption{\label{fig:higgs_EFT} One-parameter 95\% confidence exclusion regions ($-2 \log[ {\cal L}/{\cal L}_0] \gtrsim 3.84$) from precision measurements of $\sigma_{Zh}$, using results computed in the EFT. We present these as boundaries in the allowed mass parameter space for fixed representative values of $\kappa$, as annotated on each line; the \emph{unshaded} region to the lower-left of these lines is excluded for the given value of $\kappa$.  Absent the full loop computation, it is not possible to quote an error on this EFT-based result, but based on the comparison of the EFT and full-theory computations for the EWPO results, it is probable that the EFT results here are questionable in the light shaded region, $M_F/4 \lesssim m_\chi \lesssim M_F/2$, and are almost certainty invalid in the dark-shaded region, $m_\chi \gtrsim M_F/2$, where the results have consequently been masked. See \tabref{sensitivities} for the assumed sensitivities of the experiments.} 
\end{figure*}

In  \figref{higgs_EFT} we show the regions in the mass parameter space ($m_\chi$, $M_F$) which, for the given fixed value of $\kappa$, would yield a value of $\sigma_{Zh}$ in conflict with the projected 95\% confidence limits as given in \tabref{sensitivities}. We can see that the best sensitivity comes from the CEPC and FCC-ee. At the CEPC, the lower bounds on $M_F$ range from 590~GeV to 7~TeV for $\kappa \sim 1.0- 4.0$\footnote{$\kappa = 4.0$ is a fairly large coupling; we display these results only with the explicit point of indicating that a very large coupling is required to probe this region of parameter space.}  in the limit where $m_\chi \ll M_F$. These lower limits rise to approximately 660~GeV to 10~TeV for $m_\chi \sim M_F / 4$, which is roughly where we begin questioning the validity of the EFT. (The limits on $M_F$ in the region where the EFT results are valid are more sensitive to the value of $m_\chi$ than the electroweak precision limits.) 
The FCC-ee constraints are again a little stronger than CEPC constraints. Lower limits on $M_F$ in the limit where $m_\chi \ll M_F$ range from approximately 630~GeV to 7.8~TeV for $\kappa \sim 1.0-4.0$.

The constraints obtained at the ILC are somewhat weaker in all three of the considered scenarios than either the CEPC or FCC-ee constraints. 
Note that care was taken here in combining limits where runs are at different energies as $(\Delta \sigma_{Zh} )/ \sigma_{Zh}^{\textsc{sm}}$ is energy-dependent (see comment in caption on \tabref{sensitivities}). The most optimistic ILC scenario with several ab\up{$-1$} of data yields lower limits on $M_F$ for $m_\chi \ll M_F$ that are approximately 490~GeV and 5.9~TeV, respectively, for $\kappa \sim 1.0$ and 4.0. 

In none of these cases are the limits from the precision Higgsstrahlung measurement competitive with the electroweak precision programs at these future colliders in imposing constraints on this specific model; nevertheless, these results do demonstrate that the $\sigma_{Zh}$ measurement would provide a strong complimentary constraint on closely allied models where the $T$ parameter is dialed away, as we discussed previously. 

Although a large number of operators contribute to $\sigma_{Zh}$, a partial and heuristic understanding of the generic difference in the strength of these limits can be obtained by examining a subset of the operators relevant for the generation of $T$ and $\Delta \sigma_{Zh}$. Suppose ${\cal L} = {\cal L}_{\textsc{sm}} + \frac{a}{2\Lambda^2} ( H^\dagger D_\mu H - h.c. )^2 + \frac{b}{2\Lambda^2} ( \partial_\mu |H|^2 )^2$. It can then be easily shown that $T = ( a v^2 ) / ( \alpha_e \Lambda^2 ) \sim 10^2\, a( v^2 / \Lambda^2)$, and a little more work shows that $\Delta \sigma_{Zh} / \sigma_{Zh} \approx - ( v^2 / \Lambda^2) ( b + 0.83 a )$;\footnote{The `0.83' here is the numerical value of a fairly complicated function of the gauge couplings, assuming $g = 0.648, g'=0.358$.} note that the two operators contribute almost equally to the Higgstrahlung cross-section. The anticipated one-parameter 95\% confidence measurement uncertainties on $T$ (restricted to $S=U=0$) are $2.0\times10^{-2},\, 1.5\times10^{-2},$ and $8.9\times10^{-3}$ for the CEPC baseline, ``Improved $\Gamma_Z, \sin^2\theta$'', and ``Improved $\Gamma_Z, \sin^2\theta,m_t$'' scenarios, respectively.\footnote{These are obtained from the values in \tabref{sensitivities} as $\sigma_{T, \text{1 parameter}}^{\text{95\% CL}} = 1.96\, \sigma_{\textsc{T}} \, \sqrt{ 1 - \rho_{\textsc{ST}}^2 }$.}  The resulting 95\% confidence lower bounds on $\Lambda / \sqrt{|a|}$ are approximately 20~TeV, 23~TeV and 29~TeV, respectively. On the other hand, the 95\% confidence measurement uncertainty on the Higgsstrahlung cross-section at CEPC is projected to be $\sim 1$\% (corresponding to the 68\% confidence projection $\Delta\sigma_{Zh}/\sigma_{Zh} = 0.5\%$ in \tabref{sensitivities}), which yields the limit $ \Lambda / \sqrt{ | b + 0.83 a | } \gtrsim 2.5$~TeV. It is clear that the latter bounds are significantly weaker than the EWPO constraints for roughly equally-sized Wilson coefficients, which is the scenario one would expect when these operators are generated at the same loop order (provided of course that there is no symmetry---e.g., custodial---which prevents this). Parametrically, it is clear that the relative strength of the limits can be traced directly to the enhancement of $T$ by a factor of $1/\alpha_e$.

\subsection{Loop computation of EWPO}
\label{sect:loops}

By design, the EFT we have constructed is valid only in the region of parameter space where $M_F \gg m_\chi$, and it is in this region of parameter space where this model supplies a UV completion of the fermionic Higgs portal operator $H^\dagger H \bar{\chi} \chi$. One could of course also construct other EFTs valid in the region $m_\chi,M_F \gg v \sim m_W$, but for similar particle masses, $m_\chi \sim M_F$, or for the opposite mass hierarchy, $m_\chi \gg M_F$. Given the amount of work required to obtain the results already presented, it appears to be more efficient to simply perform the full one-loop computations of $S,T$; this computation is detailed in Appendix \ref{app:STloops} and was performed in the mass-eigenstate basis of neutral fermions, accounting for mixing angle effects. The full loop computation of $\sigma_{Zh}$ is beyond the scope of this paper,  as we have already established from the EFT results that the Higgsstrahlung bounds are expected to be somewhat weaker than those from EWPO.

\begin{figure*}[t]
\centering
\includegraphics[width = 0.49\textwidth]{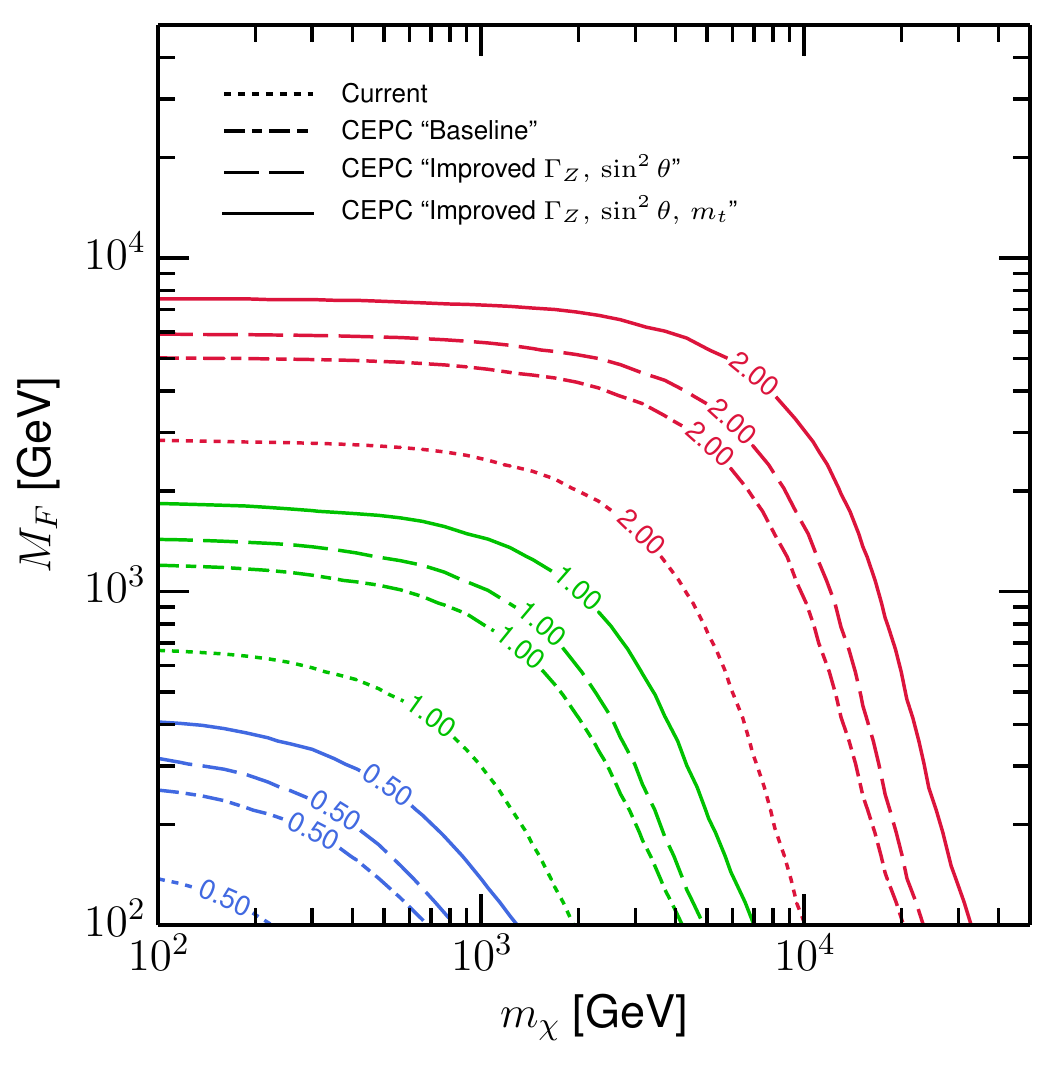}
\includegraphics[width = 0.49\textwidth]{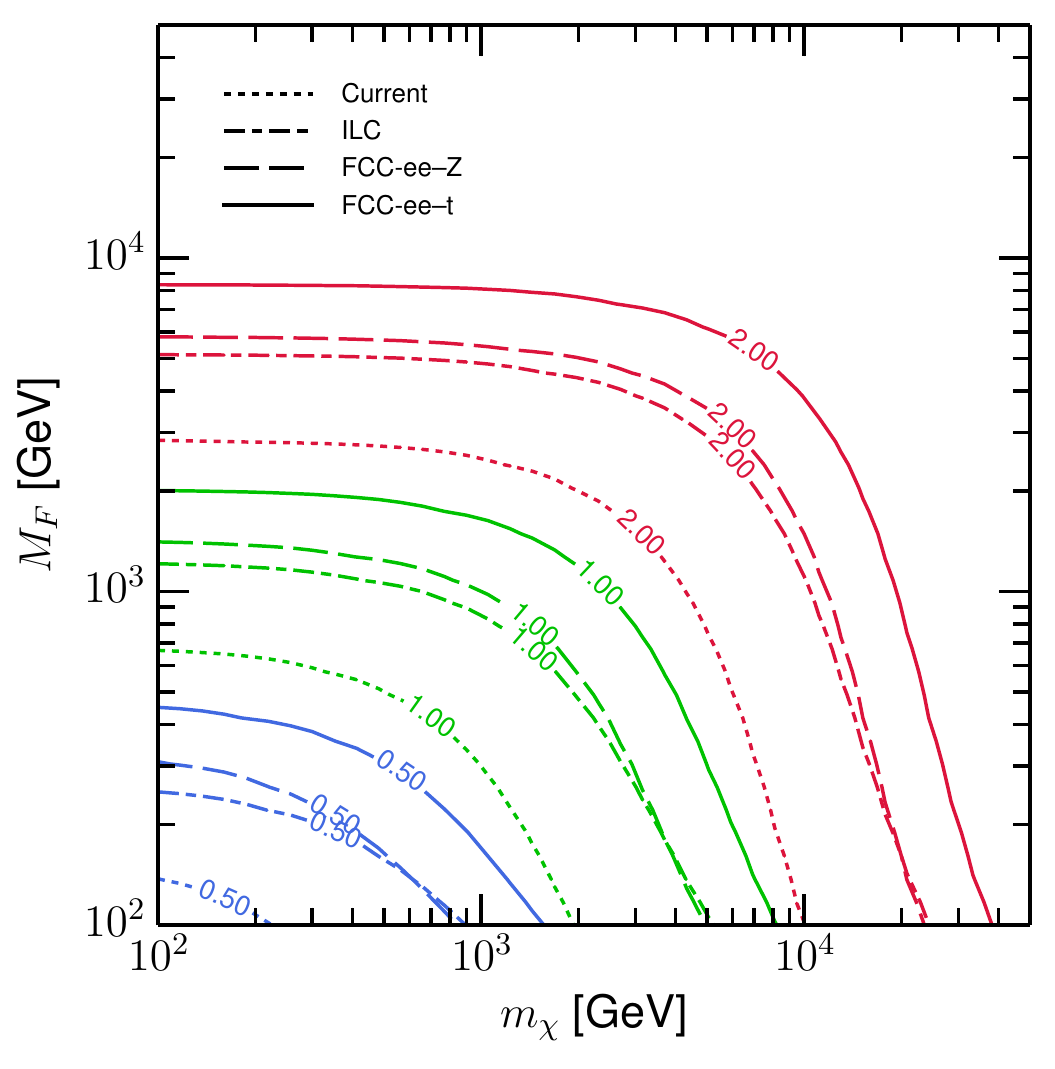}
\caption{\label{fig:ewpo_full} As for \figref{ewpo_EFT}, two-parameter 95\% confidence exclusion regions ($-2 \log[ {\cal L}/{\cal L}_0] \gtrsim 5.99$) from measurement of the $S$ and $T$ parameters, except these results are computed using the full one-loop results described in section \ref{sect:loops} and hence are valid for the full range of masses shown. Note that $m_\chi$ and $M_F$ are the input mass-parameters, and are not the physical masses of the neutral fermions; see Appendix \ref{app:STloops}. }
\end{figure*}

The exclusion regions due to precision electroweak constraints arising from the full loop computation of $S,T$ are shown in \figref{ewpo_full}. In the limit where $m_\chi \ll M_F$ it is clear that the results agree well with the EFT computation (c.f., \figref{ewpo_EFT}), the small differences being ascribable to our neglect of dimension-8-and-higher operators, as well as the running between scales, in the EFT. Note that if the full result for $S,T$ is expanded out to ${\cal O}(v^2/M_F^2)$ to find the contribution to the result ascribable to dimension-6 operators, we find all the same analytic (in the masses) pieces of the result as in the EFT computation, but naturally also find non-analytic factors $\sim \log [m_\chi^2/M_F^2]$ which are absent in our EFT result as we neglected the running of the Wilson coefficients between scales $\mu = M_F$ and $\mu = m_\chi$; however, for the contributions to $S$ and $T$, these are suppressed by at least the third power of $m_\chi/M_F$ and are thus negligibly small whenever the EFT which we constructed is valid.

As we have already discussed the results in the region $m_\chi \ll M_F$ in connection with the EFT results, we focus here on the parameter space not covered by the EFT. Of course in this region, the theory does not provide a UV completion of the fermionic Higgs portal operator $H^\dagger H \bar{\chi}\chi$, although in the limit $m_\chi \gg M_F$, it would provide a completion of the fermionic-doublet Higgs portal operator $H^\dagger H \bar{F}F$. 

The most obvious point is that a significantly larger region of the $m_\chi$ parameter space at small $M_F$ is ruled out than \emph{vice versa}. For example, the fully improved CEPC results indicate that if $M_F = 100$~GeV,  then $m_\chi < 33$~TeV can be constrained for $\kappa \sim 2.0$, compared to $M_F < 7.7$~TeV being constrained if $m_\chi = 100$~GeV. This pattern is generic for all of the experiments; it traces its origin to the fact that  for $m_\chi \ll M_F$, $S$ and $T$ are both positive and, roughly speaking, $|S| \sim 0.1 |T|$ near the exclusion limit, whereas for $M_F \ll m_\chi$, $T$ is positive and $S$ is negative, with $|S| \sim |T|$ near the exclusion limit (indeed, $|S|> |T|$ is possible here). The larger deviation from $(S,T)=(0,0)$ in this region leads to the stronger limits. There is also a stronger dependence of the boundary of the exclusion region on $M_F$ at large $m_\chi$ than \emph{vice versa}. Both of these behaviours are due to the fact that, parametrically, $S \sim \Delta M / M_F$ where $\Delta M$ is the mass-splitting in the doublet: $\Delta M \equiv M_{F}^{\text{neutral}} - M_{F}^{\text{charged}} = v (\kappa^2/ 2) [ v / (M_F-m_\chi) ]$. This parametric dependence arises because $S$ is zero in the weak-isospin-symmetric limit of the theory, and because it is always $F$ fermions which couple directly to $W$ and $Z$. This implies that the leading contributions are $S \sim \kappa^2v^2/M_F^{2}$ for $m_\chi \ll M_F$, but $S \sim  -\kappa^2v^2/(M_F m_\chi)$ for $M_F \ll m_\chi$; clearly then, at light $M_F$ and heavy $m_\chi$, $|S|$ will be larger than if the mass hierarchy were reversed.

The slightly different shapes of the ILC and FCC-ee exclusion regions when $m_\chi \gtrsim M_F$, as seen in the right plot of \figref{ewpo_full}, arise from the different alignments of the 68\% coverage likelihood contours in the $(S,T)$ plane \cite{Fan:2014vta} for the ILC compared to those for FCC-ee, which in turn is due to the different projected sensitivites to $m_W, \sin^2\theta^l_{\text{eff}}$ and $\Gamma_Z$ at each collider (see also \citeR{MishimaTalk}). With reference to figure 1 in \citeR{Fan:2014vta}, it is clear that if $S,T>0$ and $|S|\ll|T|$, as is the case for $m_\chi \ll M_F$, the limits from FCC-ee-Z (TLEP-Z) are more constraining than those for the ILC; whereas if $T>0$, and $S\sim -T$, as is the case for for $M_F \ll m_\chi$, the ILC and FCC-ee-Z limits will be about equally constraining, with the ILC actually marginally better, as is visible in the right panel of our \figref{ewpo_full} for $M_F\lesssim150$~GeV.

\section{Conclusions}
\label{sect:conclusions}
In this paper we examined two possible models which UV-complete the CP-even fermionic Higgs portal operator $H^\dagger H \bar{\chi}\chi$, and investigated how future precision electroweak and precision Higgsstrahlung measurements at the proposed ILC, FCC-ee and CEPC high-energy $e^+e^-$ colliders could constrain these models. In the first model, the ``scalar completion,'' a scalar singlet acts as a mediator between vector-like Dirac fermion $\chi$ and the SM Higgs field $H$, with a Yukawa coupling $\kappa_S S \bar{\chi}\chi$ to the $\chi$ and a coupling $a\, m_S S|H|^2$ to the Higgs doublet. In the second model, the ``fermionic completion,'' a vector-like $SU(2)$ doublet of Dirac fermions couples to the $\chi$ and $H$ field with a Yukawa-like interaction $\kappa \bar{F} H \chi + \text{h.c.}$.

For the scalar completion, we estimated that the effect of the new $\chi$ particle will be subdominant (i.e., loop-suppressed) for the computation of the electroweak precision variables $S$ and $T$ and the Higgs couplings to the SM particles, compared to the dominant effect of the singlet mediator itself. The limits which can be placed on the SM augmented with such a singlet scalar mediator have already been extensively studied in the literature (e.g., \citeR{Henning:2014gca}): a precision Higgsstrahlung cross-section measurement $(\Delta \sigma_{Zh})/\sigma_{Zh} \sim 0.5\%$ can place a 95\% confidence upper limit on $m_S/a$ around 2.5~TeV, while the precision electroweak limits are only marginally weaker (despite $S$ and $T$ only being loop-induced by running). 

Our main focus was on the fermionic completion. We constructed an EFT valid in the limit $v \lesssim m_\chi \ll M_F$ and examined the 95\% exclusion reach on the mass parameter space, for a variety of projected sensitivities for the precision electroweak and Higgsstrahlung measurements. Provided the coupling constant $\kappa$ is ${\cal O}(1)$, we found that the precision electroweak limits are very powerful and primarily driven by $T \sim (\alpha_\kappa/\alpha_e) S \gg S$, owing to the violation of custodial symmetry. Already with current limits, the 95\% confidence exclusion reach on $M_F$ for $m_\chi \ll M_F$, is up to $M_F \sim 675$~GeV (2.9~TeV) for $\kappa = 1.0\ (2.0)$. For the most optimistic projections we consider for the various possible configurations of the ILC, CEPC and FCC-ee colliders, these lower limits rise to 1.2~TeV (5.3~TeV), 1.8~TeV (7.7~TeV), and 2.0~TeV (8.5~TeV), respectively (see \figref{ewpo_EFT}).

For this model, the precision electroweak limits are generically more powerful than the precision Higgsstrahlung cross-section measurement due to the parametric enhancement of $T$ over $(\Delta \sigma_{Zh})/\sigma_{Zh}$ by $\sim 1/\alpha_e$. Nevertheless, precise measurement of the latter also yields good complementary exclusion reach, which is useful because closely-allied UV completions in which the $T$ parameter can be made small will therefore still be fairly strongly constrained by this measurement (other probes, such as $h\rightarrow \gamma\gamma$, may also become important). The most optimistic 95\% confidence exclusion lower limits from Higgsstrahlung on $M_F$, in the limit $m_\chi \ll M_F$, for the various possible configurations of the ILC, CEPC and FCC-ee colliders, are 490~GeV (1.5~TeV), 590~GeV (1.9~TeV), and 620~GeV (2.1~TeV), for $\kappa = 1.0\ (2.0)$,  respectively (see \figref{higgs_EFT}).

We also considered the full one-loop computation of the electroweak precision observables in the more general mass parameter space where this model does not necessarily provide a UV completion for the CP-even fermionic Higgs portal as we have defined it. The one-loop computation was found to agree well with the EFT computation for $m_\chi \ll M_F$. The model is significantly more constrained for the `opposite' mass hierarchy $m_\chi \gg M_F$ owing to the fact that $S$ and $T$ are comparably large in this region; for example, we find 95\% confidence lower limits on $m_\chi$ for $M_F = 100$~GeV are typically a factor of 3--5 times higher than the corresponding lower limits on $M_F$ for $m_\chi = 100$~GeV (see \figref{ewpo_full}).

Additionally for the fermionic completion, the appearance in the EFT analysis of operators modifying the coupling of the Higgs to SM fermions (see \tabref{shifted_Lag_ops}) raises the prospect of a modification to the rate for $h\rightarrow b\bar{b}$ which will be measured to fairly high accuracy at lepton colliders; we defer investigation of this point to future work.

Overall, we see that the sensitivities possible at future $e^+e^-$ machines through measurements of precision electroweak observables and the Higgsstrahlung cross-section will allow significant improvements in the exclusion reach for the CP-even fermionic Higgs portal over current limits, pushing into the (multi-)TeV range of particles masses, well beyond the direct reach of the LHC.

\acknowledgments
We would like to thank Nathaniel Craig for useful correspondence. This work was supported in part by the Kavli Institute for Cosmological Physics at the University of Chicago through grant NSF PHY-1125897 and an endowment from the Kavli Foundation and its founder Fred Kavli. L.-T.W. is supported by the DOE Early Career Award under grant de-sc0003930.

\appendix
\section{One-loop computation of $S$ and $T$}
\label{app:STloops}

\begin{figure}[b]
\centering
\caption{ \label{fig:STloop} A one-loop contribution to the $IJ$ self-energy function arising from particles $A$ and $B$ running in the loop.}
\includegraphics[width=0.45\textwidth]{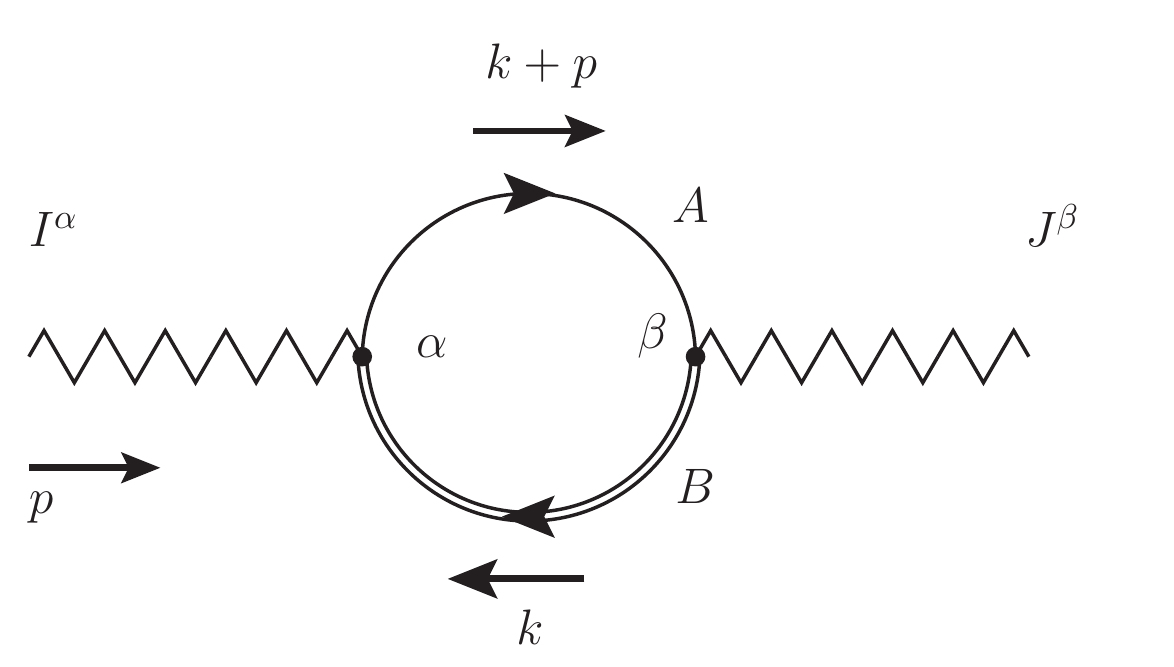}
\end{figure}

The one-loop computation of $S$ and $T$ was performed in dimensional regularization in the \MS scheme. The $IJ$ self-energy (where $I$ and $J$ are SM vector bosons) arises from diagrams such as those in \figref{STloop}. For particles $A$ and $B$ with masses $M_A$ and $M_B$, respectively, running in the loop, we find a contribution
\begin{align}
\Pi_{IJ;AB}(p^2)
&= \frac{g_{IAB}\cdot g_{JAB}}{12\pi^2} 
	\lb[ \begin{array}{l} 
   		 \text{A}_0\!\left(M_A^2\right) + \text{A}_0\!\left(M_B^2\right)  - (M_A^2 + M_B^2) + \dfrac{1}{3} p^2 - p^2 \  \text{B}_0\!\left(p^2,M_A^2,M_B^2\right)  \\[2ex]
		+ \dfrac{1}{2} \text{B}_0\!\left(p^2,M_A^2,M_B^2\right) \left( (M_A - M_B)^2 -4 M_A M_B \right)\\[2ex]
		+ \dfrac{1}{2} \dfrac{\text{B}_0\!\left(p^2,M_A^2,M_B^2\right) - \text{B}_0\!\left(0,M_A^2,M_B^2\right)}{p^2}\ \left(M_A^2-M_B^2\right)^2\\ +\ \text{counter-terms}
   	\end{array}  \rb] 
\end{align}
where $A_0$ and $B_0$ are the usual Passarino-Veltman scalar integrals \cite{Passarino:1978jh}. Expanding $\Pi(p^2)$ in powers of the gauge-boson momentum as $\Pi(p^2) = \Pi(0) + p^2\, \Pi'(0) +  \cdots$, we find, at the \MS scale $\mu$,
\begin{align}
\Pi_{IJ;AB}(0) 
&= \frac{g_{IAB}\cdot g_{JAB}}{16\pi^2} \nonumber \\[1ex]
&\quad \quad \times 
	\lb( \begin{array}{l} 
M_A^2 -4 M_A M_B+M_B^2 \\[2ex]
+ \dfrac{1}{3} \left(M_A^2-M_B^2\right)^{-1} \lb[ \begin{array}{l} 
2 M_A^2M_B^2\, \log \left( M_B^2 / M_A^2 \right) \\
- 2M_A^2\left(6 M_A M_B-3 M_A^2+M_B^2\right) \log \left( \mu ^2 / M_A^2 \right)\\
+2 M_B^2 \left(6 M_A M_B-3 M_B^2+M_A^2\right)  \log \left( \mu ^2 / M_B^2 \right) \end{array} \rb] \end{array} \rb)  \label{eq:STloopcontrib}\\
&\longrightarrow 0 \ \ \text{as} \ \ M_A \rightarrow M_B, 
\label{eq:STloopcontribSAME}
\end{align}
and
\begin{align}
\Pi'_{IJ;AB}(0)
&= \frac{g_{IAB}\cdot g_{JAB}}{16\pi^2} \times \lb(- \dfrac{2}{9} \rb) \lb[ M_A^2-M_B^2 \rb]^{-3} \nonumber \\
&\quad \quad \times
\lb( \begin{array}{l} \left(M_A^2-M_B^2\right) \left( \begin{array}{l}
										 \phantom{+} 6M_A^2 \left(M_A^2- M_B^2\right) \log \left(\mu ^2/M_A^2\right) \\ 
										 +6 M_B^2 \left(M_B^2-M_A^2\right) \log \left(\mu ^2/M_B^2\right)\\
										 +9 M_A^3 M_B-16 M_A^2 M_B^2+9 M_A M_B^3+2 M_A^4+2 M_B^4 \end{array} \right) \\
   +6 M_A^2 M_B^2 \left(-3 M_A M_B+M_A^2+M_B^2\right) \log \left(M_A^2/M_B^2\right) \end{array}\rb)\label{eq:STloopcontrib2} \\[2ex]
&\longrightarrow \frac{g_{IAB}\cdot g_{JAB}}{16\pi^2} \times \lb( -\dfrac{4}{3} \rb) \log \left(\mu ^2/M^2\right) \ \ \text{as}\ \ M_A \rightarrow M_B \equiv M.\label{eq:STloopcontrib2SAME}
\end{align}

In the broken phase of the theory, the mass-eigenstate basis of new fermions is $\{ n_1, n_2, C \}$, where  
\begin{align}
\begin{pmatrix} n_1 \\ n_2 \end{pmatrix} &= \begin{pmatrix} \cos \theta & - \sin \theta \\ \sin \theta & \cos\theta \end{pmatrix} \begin{pmatrix} \chi \\ N \end{pmatrix}. \label{eq:mass_rotation}
\end{align}
Here the mixing angle is given by
\begin{align}
\tan2\theta = \frac{ \sqrt{2} \kappa v}{ M_F - m_\chi },\label{eq:theta_defn}
\end{align}
and these particles have eigenmasses
\begin{align}
M_1 &= \frac{1}{2} \lb[ m_\chi + M_F - \bigg[ (M_F - m_\chi)^2 + 2\kappa^2 v^2 \bigg]^{1/2} \rb], \label{eq:M1_mass}\\
M_2 &= \frac{1}{2} \lb[ m_\chi + M_F +\bigg[ (M_F - m_\chi)^2 + 2\kappa^2 v^2 \bigg]^{1/2} \rb], \quad \text{and} \quad M_C = M_F.
\end{align}
The $W^+W^-$, $ZZ$, $Z\gamma$ and $\gamma\gamma$ self-energy functions and their derivatives can be immediately found by summing the contributions from eqs.~(\ref{eq:STloopcontrib})--(\ref{eq:STloopcontrib2SAME}) over all the allowed new-physics particles which run in the loop (see \tabref{LoopParticles}), using the following results for the couplings:
\begin{align}
g_{ZCC} &= \frac{g^2-(g')^2}{2\sqrt{g^2+(g')^2}}, 		&	g_{\gamma CC}&= \frac{gg'}{\sqrt{g^2+(g')^2}}, 			\label{eq:g1}\\
g_{Z11} &= - \frac{\sqrt{g^2+(g')^2}}{2}  \sin^2\theta,		&	g_{W^+C1} &= g_{W^-1C} = - \frac{g}{\sqrt{2}} \sin\theta,	\label{eq:g2}\\
g_{Z22} &= - \frac{\sqrt{g^2+(g')^2}}{2}  \cos^2\theta,		&	g_{W^+C2} &= g_{W^-2C} = \frac{g}{\sqrt{2}} \cos\theta, 	\label{eq:g3}\\
\text{and}\ \ g_{Z12} = g_{Z21} &= \frac{\sqrt{g^2+(g')^2}}{2}  \sin\theta \cos\theta. \label{eq:g4}
\end{align}

\begin{table}
\centering
\caption{ \label{tab:LoopParticles} The new-physics particles $(A,B)$ which run in the self-energy loop shown in \figref{STloop}.}
\begin{tabular}{ll}
\textbf{Self-energy $\bm{IJ}$}	&	\textbf{Particles $\bm{(A,B)}$ in loop} \\ \hline
$W^+W^-$		&	$(C,1)$, $(C,2)$						\\
$ZZ$				&	(1,1), (2,2), (1,2), (2,1), $(C,\overline{C})$		\\
$Z\gamma$		&	$(C,\overline{C})$						\\
$\gamma\gamma$	&	$(C,\overline{C})$						
\end{tabular}
\end{table}

$S$ and $T$ can then be computed from their definitions, eqs.~(\ref{eq:deriv_S}) and (\ref{eq:2}), and the results shown here; we have checked explicitly that all terms depending on $\mu$ cancel in the results for $S$ and $T$, as they should. Note that $T$ arises as a small difference between two numerically large terms, and so care is required to obtain a numerically accurate result if these terms are not manually cancelled.

\section{A clarification on matching in the presence of mixing}
\label{app:mix_match}
This Appendix aims to clarify a point about the matching at the scale $\mu = M_F$ which may not be immediately familiar to the reader: when $\langle H \rangle \neq 0$, the mass-eigenstate $\chi$ field we write in the intermediate-energy EFT\footnote{In this Appendix, `EFT' always refers to the intermediate-energy EFT, after the $F$ field is integrated out at $\mu = M_F$, but before the $\chi$ field is integrated out at $\mu = m_\chi$. This is the EFT described by the operators and coefficients in Tables \ref{tab:high_scale_operators} and \ref{tab:high_scale_coefficients}.} should \emph{not} be thought to be identified with the (mass-eigenstate admixture) $\chi$ field in the UV theory; rather the $\chi$ field in the EFT should instead be identified with the light-mass-eigenstate $n_1$ in the UV theory. This fact notwithstanding, it is common practice, although something of an abuse of notation, to still call the field in the EFT `$\chi$'.

Consider first a general argument. We have presented both the EFT and UV theory in terms of $SU(2) \times U(1)_Y$--symmetric operators. However, because (a) the limit $\langle H \rangle \rightarrow 0$ is smooth in the sense that the $F$ doublet in the UV theory remains sufficiently massive $(M_F \gg \kappa v,\, m_\chi)$ to be integrated out, and that the EFT power-counting is in $\kappa \langle H \rangle / M_F$, so no EFT operators or Wilson coefficients diverge as $\langle H \rangle\rightarrow 0$; (b) the only source of EWSB in either the UV theory or the EFT is the existence of a non-zero $\langle H \rangle$; and (c) the EFT is explicitly constructed to match the amplitudes of the UV theory at $\mu = M_F$, it follows that when $\langle H \rangle \neq 0$, the EFT must \emph{by construction} capture all the $\langle H \rangle$-dependent effects which are present in the UV theory, at each order in the EFT power counting. 

Put another way, up to the truncation order of the EFT, one obtains the same physical predictions from the EFT by (a) matching in the broken phase ($\langle H \rangle\neq 0$) of the UV-theory by integrating out the heavy-mass-eigenstates $n_2$ and $C$, and writing the EFT in terms of the light-eigenstate $n_1$ and the physical Higgs field $h$; or (b) matching in the unbroken phase ($\langle H \rangle = 0$), integrating out the $F$ doublet whole and writing the EFT in terms of (the EFT field) $\chi$ and the $SU(2)$-doublet field $H$, and then turning on a non-zero $\langle H \rangle$ in the EFT after the matching. Once $\langle H \rangle$ is turned on, the EFT automatically includes the corrections to the $\chi$ field which allow its identification with the same physical degree of freedom represented by the light-mass-eigenstate $n_1$ of the UV theory. 

We emphasize that it is by construction that the EFT must reproduce the UV-theory amplitudes and so must capture the mixing of the fermions $\chi$ and $N$ in the UV-theory. In the $SU(2)$-symmetric EFT, this mixing is present in the form of the higher-dimension operators of the EFT. We now show examples of this by explicit computation.

\subsection{Example 1: two-point functions of the UV-theory $n_1$ field and the EFT field $\chi$}
The clearest demonstration that the EFT field $\chi$ represents the same physical degree of freedom as the UV-theory field $n_1$ is to show that these two fields have the same two-point function; or, equivalently, that when they both have canonical kinetic terms in their respective Lagrangians, they have the same physical mass (up to corrections at the truncated order in the EFT power-counting). 

The UV-theory $n_1$ field has a canonical kinetic term and mass given in  \eqref{M1_mass}. Expanding this result out for $M_F \gg \kappa v, m_\chi$ gives
\begin{align}
M_1 \approx m_\chi - \frac{\kappa^2v^2}{2M_F} - \frac{m_\chi \kappa^2 v^2 }{2M_F^2} + \frac{\kappa^4 v^4 - 2 m_\chi^2 \kappa^2 v^2}{4 M_F^3} + \mathcal{O}(M_F^{-4}).
\label{eq:M1_expansion}
\end{align}

The EFT $\chi$ field does not have canonical kinetic terms. The relevant higher-dimensional operators in the EFT are (see Tables \ref{tab:high_scale_operators} and \ref{tab:high_scale_coefficients}) 
\begin{align}
\mathcal{L}_{\text{EFT}} &\supset i \bar{\chi} \slashed{\partial} \chi - m_\chi \bar{\chi}\chi +  \frac{c_5}{M_F}\, \mathcal{O}_5 + \frac{c_{6B}}{ M_F^{2}}\, \mathcal{O}_{6B} + \frac{c_{7A}}{ M_F^{3}}\,  \mathcal{O}_{7A}  + \mathcal{O}(M_F^{-4}) \\
&= i \bar{\chi} \slashed{\partial} \chi - m_\chi \bar{\chi}\chi  +  \frac{\kappa^2}{M_F} (H^\dagger H) \bar{\chi}\chi + \frac{1}{2} \frac{\kappa^2}{M_F^2} (H^\dagger H) i ( \bar{\chi} \slashed{\partial} \chi - \text{h.c.} ) \nonumber  \\
&\qquad  - \frac{1}{2} \frac{\kappa^2}{M_F^3} (H^\dagger H) ( \bar{\chi} \Box \chi + \text{h.c.} ) + \mathcal{O}(M_F^{-4}).
\end{align}
Breaking electroweak symmetry by setting $\langle H \rangle = ( 0 , v / \sqrt{2} )^T$, the relevant terms become
\begin{align}
\mathcal{L}_{\text{EFT}} &\supset i \bar{\chi}\slashed{\partial}\chi \lb[ 1 + \frac{\kappa^2v^2}{2M_F^2} \rb] - \bar{\chi}{\chi} \lb[ m_\chi - \frac{\kappa^2v^2}{2M_F} \rb] 
- \frac{\kappa^2v^2}{4M_F^3} \lb[ \bar{\chi} \stackrel{\leftarrow}{\Box} \chi +  \bar{\chi} \stackrel{\rightarrow}{\Box} \chi \rb] + \mathcal{O}(M_F^{-4}),
\end{align}
This can be re-cast to canonical form by performing a field-redefinition, which does not change the physical content of the theory:
\begin{align}
\chi \rightarrow \lb[ 1 - \frac{\kappa^2 v^2}{4M_F^2} - \frac{ m_\chi \kappa^2 v^2}{4M_F^3} \rb] \chi - \frac{\kappa^2v^2}{4M_F^2} \frac{i\, \slashed{\partial}\chi}{M_F} + \mathcal{O}(M_F^{-4}).
\end{align}
In terms of the redefined field, we have
\begin{align}
\mathcal{L}_{\text{EFT}} &\supset i\bar{\chi}\slashed{\partial}\chi - \bar{\chi}{\chi} \lb[ m_\chi - \frac{\kappa^2v^2}{2M_F} - \frac{m_\chi\kappa^2v^2}{2M_F^2} + \frac{\kappa^4v^4-2m_\chi^2 \kappa^2v^2}{4M_F^3} \rb] + \mathcal{O}(M_F^{-4}).
\label{eq:chi_mass_expansion}
\end{align}
Comparing eqs.\ (\ref{eq:M1_expansion}) and (\ref{eq:chi_mass_expansion}), we have established the desired result.

\subsection{Example 2: $Z$-boson coupling to the EFT field $\chi$}
We now look at an example for the EFT $\chi$ interaction terms. Consider that the UV-theory $\chi$ field is an exact SM-singlet, while $N\subset F$ has SM gauge-couplings. When $\langle H \rangle\neq0$, the mass-basis rotation/mixing, \eqref{mass_rotation}, generates SM gauge-couplings for both the $n_{1,2}$ mass-eigenstate fermions; see eqs.\ (\ref{eq:g1}) to (\ref{eq:g4}). 

If the EFT field $\chi$ were to be identified exactly with the $\chi$ field in the UV theory, it would necessarily follow that the former should have no SM gauge-couplings; if instead the EFT field $\chi$ were identified with the light-mass-eigenstate $n_1$ of the UV theory, we would expect the former to couple to, e.g., the $Z$-boson. The second alternative is the correct one, as the following short argument shows.

Consider that in the UV theory the coupling constant of the $n_1$ vector current to the $Z$-boson is $g_{Z11} = - \frac{1}{2} \sqrt{g^2+(g')^2} \sin^2 \theta$ where $\tan 2 \theta = \sqrt{2} \kappa v / (M_F-m_\chi)$; see eqs.\ (\ref{eq:theta_defn}) and (\ref{eq:g2}). In the limit where $M_F \gg \kappa v ,\, m_\chi$, we have
\begin{align}
g_{Z11} \approx  - \frac{1}{4} \sqrt{g^2+(g')^2} \; \frac{\kappa^2 v^2 }{M_F^{2}} + \cdots.
\end{align}
Meanwhile, in the EFT the following operator is present: 
\begin{align}
\mathcal{L}_{\text{EFT}} \supset \frac{c_{6A}}{M_F^{2}}\, \mathcal{O}_{6A}  \supset - \frac{1}{4} \sqrt{g^2+(g')^2} \; \frac{\kappa^2 v^2 }{M_F^{2}}\,\bar{\chi}\gamma_\mu \chi \, Z^\mu + \cdots.
\end{align}
(Note: the field-redefinition in Example 1 above does not modify this leading-order coupling.) One can readily see that the leading-order coupling of the $Z$-boson with the vector-current of the EFT $\chi$ field has the same sign and magnitude, and appears at the same order in the EFT power-counting, as the leading-order coupling of the $Z$-boson with the vector-current of the UV-theory $n_1$ field. Again, we draw the reader's attention to the fact that it is a higher-dimension operator in the EFT that has captured an effect of UV-theory fermion mixing when $\langle H \rangle\neq0$.

\bibliographystyle{JHEP}
\bibliography{flw}
\end{document}